\documentclass[12pt]{article}
\usepackage{amsfonts}
\usepackage[dvips]{graphics}
\newcommand{\Rl}{\mathbb{R}}

\newcommand{\Ir}{\mathbb{Z}}
\newcommand{\Cx}{\mathbb{C}}
\newcommand{\A}{\mathcal{A}}

\newtheorem{theorem}{Theorem}[section]
          
\newtheorem{lemma}[theorem]{Lemma}              

\newtheorem{corollary}[theorem]{Corollary}

\newcommand{\rem}[1]{{\bf Remark:}}
\newcommand{\condmat}[1]{archived as {\tt cond-mat/#1}}
\newcommand{\Section}[1]{\setcounter{equation}{0}\section{#1}}

\newcommand{\eq}[1]{(\ref{#1})}
\newenvironment{proof}{\noindent {\bf Proof: }}{\QED\medskip}
\def\QED{{\hspace*{\fill}{\vrule height .5ex width 1ex }\quad} 
    \vskip 0pt plus20pt}
\newcommand{\be}{\begin{equation}}
\newcommand{\ee}{\end{equation}}
\newcommand{\bea}{\begin{eqnarray}}
\newcommand{\eea}{\end{eqnarray}}
\newcommand{\beann}{\begin{eqnarray*}}
\newcommand{\eeann}{\end{eqnarray*}}

\newcommand{\ket}[1]{\left\vert #1\right\rangle}
\newcommand{\bra}[1]{\left\langle #1\right\vert}

\newcommand{\Prob}{{\rm Prob}}

\DeclareMathAlphabet{\mathol}{OT1}{cmr}{l}{ol}

\newcommand{\up}{\ket{\uparrow}}
\newcommand{\down}{\ket{\downarrow}}
\newcommand{\uu}{\ket{\uparrow \uparrow}}
\newcommand{\ud}{\ket{\uparrow \downarrow}}
\newcommand{\du}{\ket{\downarrow \uparrow}}
\newcommand{\dd}{\ket{\downarrow \downarrow}}

\newcommand{\alphavec}{\mathbf{\alpha}}

\newcommand{\Z}{\mathbb{Z}}
\newcommand{\ZGC}[2]{Z^{GC}(#1,#2)}
\newcommand{\floor}[1]{\lfloor{#1}\rfloor}
\newcommand{\ip}[2]{\langle{#1|#2}\rangle}
\newcommand{\op}[3]{\bra{#1} #2 \ket{#3}}

\newcommand{\avg}[2]{\langle{#1}\rangle_{#2}}
\newcommand{\avgGC}[2]{\langle{#1}\rangle^{GC}_{#2}}
\newcommand{\unity}{1\hskip -3pt \rm{I}}
\newcommand{\supp}{{\rm supp}}
\newcommand{\Num}{\mathsf{N}}
\newcommand{\navg}{\avgGC{\Num}{\Sigma,\mu}}
\newcommand{\Sc}{\mathcal{S}}
\begin{document}
{\baselineskip=10pt \thispagestyle{empty} {{\small Preprint UC Davis Math
1999-24}
%{\tt cond-mat/9908018n} and {\tt mp\_arc 99-nnn} 
\hspace{\fill}}

\vspace{20pt}

\begin{center}
{\LARGE \bf Finite-volume excitations of the 111\\
interface in the quantum XXZ model\\[27pt]}
{\large \bf Oscar Bolina, Pierluigi Contucci, Bruno Nachtergaele 
and Shannon Starr\\[10pt]}
{\large  Department of Mathematics\\
University of California, Davis\\
Davis, CA 95616-8633, USA\\[15pt]}
{\normalsize bolina@math.ucdavis.edu, contucci@math.ucdavis.edu, 
bxn@math.ucdavis.edu, sstarr@math.ucdavis.edu}\\[30pt]
%(date)\\[30pt]
\end{center}

\noindent
{\bf Abstract}
We show that the ground states of the three-dimensional XXZ Heisenberg
ferromagnet with a 111 interface have excitations localized in a
subvolume of linear size $R$ with energies bounded by $O(1/R^2)$. As
part of the proof we show the equivalence of ensembles for the 111
interface states in the following sense: In the thermodynamic limit the
states with fixed magnetization yield the same expectation values for
gauge invariant local observables as a suitable grand canonical state
with fluctuating magnetization. Here, gauge invariant means commuting with
the total third component of the spin, which is a conserved quantity of
the Hamiltonian. As a corollary of equivalence of ensembles we also prove
the convergence of the thermodynamic limit of sequences of canonical
states (i.e., with fixed magnetization).
\vspace{8pt}
\newline
{\small \bf MCS numbers:} 82B10, 82B24, 82D40 
\vfill
\hrule width2truein \smallskip {\baselineskip=10pt \noindent Copyright
\copyright\ 1999 by the authors. Reproduction of this article in its entirety, 
by any means, is permitted for non-commercial purposes.\par }}

\newpage

\Section{Introduction and main results}
\label{sec:intro}

A determining factor in the stability of the magnetic state of small 
ferromagnetic particles is the structure of the spectrum of their 
low-lying excitations. Stability against thermal (and quantum) fluctuations
is a major concern when one is interested in increasing the 
density of information stored on magnetic hard disks. Higher density of
information requires smaller magnetic particles to store the bits. The
smaller these particles get, the less stable their magnetic state tends to
be. It is also well-known that ferromagnets spontaneously form domains with 
different orientations of the magnetization. These two facts motivate us
to study the excitation spectrum of finite size ferromagnets with a domain
wall or {\it interface}. From examples, it is known that the presence of an 
interface, in general, has an effect on the low-lying excitation spectrum 
\cite{KN2,KN3}.

We consider the spin 1/2 XXZ Heisenberg model on the three-dimensional
lattice $\Ir^3$. For any finite volume $\Lambda\subset\Ir^3$, the 
Hamiltonian is given by 
\be
H_\Lambda = - \sum_{x,y\in\Lambda\atop \vert x-y\vert=1}
\Delta^{-1} (S_x^{(1)} S_y^{(1)} + S_x^{(2)} S_y^{(2)}) 
+ S_x^{(3)} S_y^{(3)},
\ee
where $\Delta>1$ is the anisotropy. It will be convenient to work with
the usual parametrization $\Delta=(q+q^{-1})/2$, $0<q<1$. Note that
in the limit $\Delta\to\infty$ ($q\to 0$), one recovers the Ising model.
The case $\Delta=1$ ($q=1$) is the XXX Heisenberg model. 

It is well-known that this model has two ferromagnetically ordered
translation invariant ground states. What is less well-known is that
there are also ground states describing an interface between two domains
with opposite magnetization. The 100 interfaces are similar to the
Dobrushin interfaces found in the Ising model. They exist for
sufficiently small temperatures, as was recently proved in \cite{BCF}.
Unlike the Ising model, the XXZ model also possesses ground states with a
rigid 111 interface at zero temperature \cite{KN2}. Its stability at
positive temperatures is still an open problem. 

In this paper we are interested in estimating the low-lying excitation\~s
above the ground state with a 111 interface. It is easy to show that the
excitation spectrum above the translation invariant ground states has a 
non-vanishing gap. In \cite{KN2} it was proved that, in the corresponding
two-dimensional model, the excitations above the 11 interface are gapless.
By an extension of the methods in \cite{LPW}, Matsui \cite{Mat2} showed
that the excitation spectrum has to be gapless in all dimensions $\geq 2$.
Here, we are interested in the nature of the low-lying excitations for the
three-dimensional model, and in particular their dependence on size.
We prove the following bound for the energy of an excitation localized in 
a finite domain $\Lambda_R$ of linear size $R$. \\

{\bf Main Result:}
{\em Excitations localized in $\Lambda_R$ have a gap $\gamma_R$ bounded by
\be
\gamma_R\leq 100\frac{q^{2(1-\delta(q,\nu))}}{(1-q^2)}\frac{1}{R^2},      
\quad \textrm{for}\quad R>70.
\label{main}
\ee
where $\delta(q,\nu)$ is an exponent between $0$ and $1/2$ that depends on
the filling factor $\nu$ of the interface plane (see explanation below),
as well as the parameter $q$.}\\

The meaning of this bound is the following. We consider the model in a finite
volume $\Lambda$, with a fixed magnetization and boundary conditions that
induce an interface.  By perturbing the ground state in a cylindrical
subvolume $\Lambda_R$, with circular cross-section of radius $R$, we then 
construct an orthogonal  state with the same magnetization. The bound 
\eq{main} is an upper bound for the difference in
energy of this state with respect to the ground state  in the limit $\Lambda
\nearrow \Ir^3$. For finite volumes $\Lambda$, the same bound holds as long as
$\Lambda$ is substantially larger than $R$.  When $R$ and the finite volume are
comparable in size, a similar bound holds but with a larger constant factor and
additional error terms (see Section 4).

The dependence on $q$ of the bound \eq{main} has some interesting features,
which we explain next. First, in the limit $q\to 1$, the bound diverges.
This means that our Ansatz for the excitations of the 111 interface does not
work for the isotropic model. This is not surprising as the isotropic model
does not have a rigid 111 interface, although it does possess gapless 
excitations, as is well-known from spinwave theory. In the limit $q\to 0$,
the Ising limit, the bound vanishes. This is to be expected, as the 111
interface contours of the Ising model are highly degenerate. 

In order to explain the role of the exponent $\delta(q,\nu)$ in \eq{main}
we first need to discuss some properties of the interface states 
themselves. For $0<q<1$,
the model has a two-parameter family of pure ground states with an interface
in the 111 direction. One parameter is an angle, playing the same role as
the angles $\phi_x$ in the Ansatz \eq{ansatz} for the excitations. The
second parameter, which is relevant for the present discussion, corresponds
to the mean position of the interface in the lattice. 
If we think of spin up at any site as describing an empty site, and spin down 
as a site occupied by a particle, the third component of the spin becomes
equivalent to the number of particles. In Section 2,
\eq{grand_canonical}, we will introduce the chemical potential $\mu$ 
to control the expected number of particles, alias the third component of 
the total spin. In the limit $q\to 0$, the filling factor $\nu$ of the 
interface has a simple interpretation: $\nu=0$ means that interface separates 
a region entirely filled with particles from a region that is empty. A non-zero $\nu$ means that there is a partially
filled plane in between the filled and the empty region, with filling factor
$\nu$. It turns out that the exponent $\delta(q,\nu)$, can be considered as  
a function of $\mu$ alone. For each value of $\mu\in\Rl$, we get an interface 
state, and $\delta$ is the distance of $\mu$ to the integers, i.e.,
$\delta(\mu)=\min(\vert\mu-\floor{\mu}\vert, \vert 1-\mu+\floor{\mu}\vert)$,
where $\floor{\mu}$ is the integer part of $\mu$. In general, the relation
between $\mu$ and $\nu$ depends nontrivially on $q$. But for all $q$,
$0<q<1$, one has $\delta(q,1/2)=0$ and $\delta(q,0)=1/2$. For further
details on the interdependence of the parameters $q,\delta,\mu$, and $\nu$,
we refer to Section 6.1.

We believe that $O(1/R^2)$ is the true behavior of the low-lying excitations.
There are indications in the physics literature that this should indeed be
the case \cite{HN}. Our rigorous bounds are obtained using the variational 
principle: If $\psi_0$ is a ground state of $H_\Lambda$, and $\psi$ is any 
other state that is linearly independent of $\psi_0$, then
\be
\gamma:=E_1-E_0\leq \frac{ \op{\psi}{H^{(q)}_{\Lambda}}{\psi} }{\|\psi\|^2}
	\cdot \frac{1}{1 
	- \frac{| \ip{\psi_0}{\psi} |^2}{\|\psi_0\|^2\|\psi\|^2}} \; .
\label{vp}
\ee
The first factor in the RHS is the energy of the perturbed state $\psi$.
The second factor is necessary to correct for the non-orthogonality of
$\psi$ and the ground state. In general, one would need to consider the
orthogonal complement of $\psi$ to the entire ground state subspace of
$H_\Lambda$. In the present case however, we know that for each eigenvalue
of the third component of the total spin, $J^{(3)}$, there is exactly one
ground state.  As we will only consider perturbations that commute with
$J^{(3)}$, it is sufficient to take the orthogonal complement of $\psi$ to
$\psi_0$.

Our ansatz for $\psi$ is of the following form
\be
\psi=\prod_{x\in\Lambda_R}e^{i2\phi_x S_x^{(3)}}\psi_0\quad .
\label{ansatz}
\ee
The energy of such a state can be written as follows
\be
\frac{\langle \psi \mid H_{\Lambda} \mid \psi \rangle}{\Vert\psi\Vert^2}
=\sum_{x\in\Lambda_R,y\in \Lambda\atop \vert x-y\vert=1}
P_{x,y}[1 - \cos(\phi_x - \phi_y)].
\label{ene_form}\ee
where the $P_{x,y}$ are probabilities determined by the interface ground
state. $P_{x,y}$ can be interpreted as the probability that the bond
$(x,y)$ belongs to ``the interface contour'', i.e., one of the sites is
occupied by an up spin and one by a down spin. These probabilities decay
exponentially fast as a function of the distance to the expected location
of the interface. In particular, this shows that the interface is rigid
and that the problem of calculating its excitation energies is quasi
two-dimensional. In fact, the next step in our proof makes this explicit. 
We consider excitations of the form \eq{ansatz} with
$$
\phi_x = \mathcal{S}\phi(\frac{x_\perp}{R}), \quad R\geq 1
$$
where $\mathcal{S}$ is a suitable scale factor, $\phi$ is a smooth function 
with compact support in $\Rl^2$, and $x_\perp$ is the component of 
$x\in\Ir^2$, orthogonal to the $111$ direction. It is shown that the energy 
$\gamma_R$ of  such excitations satisfies the bound
$$
\gamma_R\leq \frac{C(q)}{R^2}\frac{\Vert \nabla \phi\Vert_{L^2}^2}{\Vert\phi
\Vert^2_{L^2}}\quad .
$$
In principle, $\phi$ is a map from $\Rl^2$ to the circle, and as such could 
have nontrivial topology. As we will only be considering small perturbations,
this will be of no relevance here. It is, therefore, natural to take for $\phi$
an eigenfunction belonging  to the smallest eigenvalue of $-\Delta$ on a
circular domain with Dirichlet boundary conditions, which minimizes of the
Rayleigh quotient on the  RHS, i.e., the Bessel function $J_0$. This is
different from the so-called superinstanton Ansatz of Patrascioiu and Seiler in
\cite{PS}, where they use the fundamental solution of the Laplace equation,
instead of an eigenfunction.

All our results are for ground states that are eigenstates of the third
component of the total spin, which is a conserved quantity, and for 
thermodynamic limits of such states. We will call this {\em the canonical 
ensemble}. Our derivation, however, relies on an equivalence of ensembles 
result for the interface ground states of the XXZ model.  The state of the
``small'' volume $\Lambda_R$, immersed in the much larger  volume $\Lambda$, is
well approximated by a grand canonical state with  suitable chemical potential
(see Chapter 2 for the precise definitions), which does not have a fixed
magnetization. As expected, this equivalence of  ensembles holds only for
observables that commute with the third component of the total spin which are
analogous to the gauge invariant observables in  particle systems. This
equivalence of ensembles result is non-trivial. Although we only give the proof
in dimensions 3, it is straightforward to generalize the proof to all
dimensions $\geq 3$. Equivalence of ensembles (in the above sense) does not
hold for the one-dimensional model. This can be derived from the results in
\cite{GW}. In two dimensions, our method without modifications, yields the
equivalence of ensembles for volumes that  grow as $\sqrt{L}$ in the $11$
direction and as $L$ in the direction of the  interface. With additional work
one can obtain equivalence of ensembles result  for standard sequences of
increasing volumes.

As another application of equivalence of ensembles we prove the existence 
of the thermodynamic limit of sequences canonical ground states with a given
density, i.e., magnetization per site.

Concerning the gap above diagonal interface states in dimensions other than
three we can make the following comments. First of all, diagonal interface
states exist in all dimensions \cite{ASW}. In one dimension there is a spectral
gap above the ground states \cite{KN1}. In two dimensions an upper bound of
order $1/R$ was proved in \cite{KN2}. The method of this paper can be used to
obtain a bound of order $1/R^2$ also in two dimensions. In all dimensions
greater than three our method can be applied without change to obtain
equivalence of ensembles, the existence of the thermodynamic limit and an upper
bound of order $1/R^2$ for the  excitation energies.

The paper is organized as follows. Chapter 2 introduces the model and the
geometrical setting.  Chapter 3 deals with the equivalence of ensembles result
which is a main ingredient of our proofs. The bound on the excitation energy is
a product of two factors as in  \eq{vp}. A bound on the first factor, called
the {\em energy bound}, is  derived in Section 4. The second factor requires an
estimate for the inner  product of the ground state with the perturbed state,
which is derived in  Section 5. In Section 6 we prove a number of results
for the grand canonical ensemble in one dimension that we use 
in the paper. 

\Section{Interface states of the XXZ model}

\begin{figure}[t]
\begin{center}
\resizebox{!}{7truecm}{\includegraphics{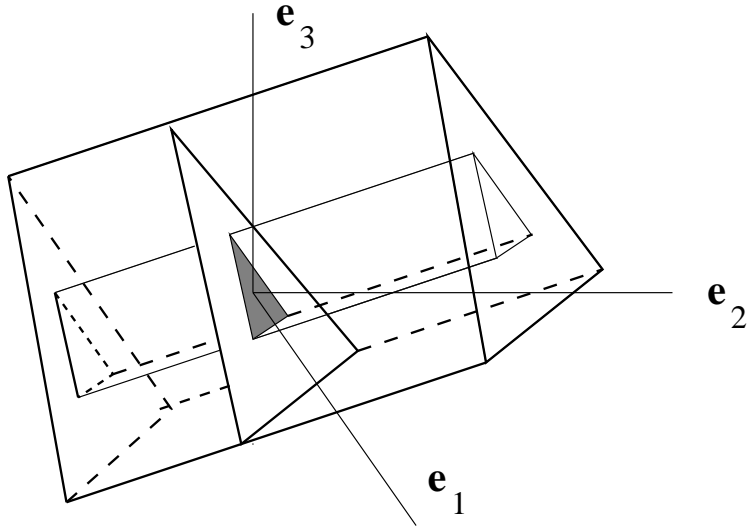}}
\parbox{14truecm}{\caption{\baselineskip=16 pt\small\label{fig:cylinder}
Example of a cylindrical $\Lambda$ embedded in $\Ir^3$.
A small cylindrical subvolume as used in the construction of the perturbed
states is also shown.}
}
\end{center}
\end{figure}

Our magnet occupies a volume $\Lambda$ which is a subset of $\Ir^3$.
Let $e_1,e_2,e_3$ denote the standard basis vectors in $\Ir^3$.
(See Figure \ref{fig:cylinder}.)
We let $l(x)$ denote the signed distance from the origin: 
$l(x) = x^1 + x^2 + x^3$, where $x = (x^1, x^2, x^3)\in\Ir^3$. 
Then
\be
B(\Lambda)=\{(x_0,x_1) :  | x_0 - x_1 | = 1, l(x_1)=l(x_0)+1\}
\label{set_bonds}\ee
describes the set of oriented bonds in $\Ir^3$. The infinite {\it stick} 
$\Sigma_0^\infty$ is, by definition, the set of vertices of the form
$$
\ldots -e_2-e_3,-e_3,0,e_1,e_1+e_2,e_1+e_2+e_3,e_1+e_2+e_3+e_1,\ldots
$$
For any even integer $L$, the finite stick $\Sigma_0$ of length $L+1$ is 
then given by
$$
\Sigma_0=\{x\in\Sigma_0^\infty\mid -L/2\leq l(x)\leq L/2\}\quad.
$$
We will take for $\Lambda$ is a cylindrical region whose axis points in the 
111 direction, where by {\it cylindrical\/} we mean that $\Lambda$ can be 
obtained from a subset $\Gamma$ of the $l(x)=0$ plane, which we will call the 
base, by adding to all vertices $x\in\Gamma$ the finite stick $\Sigma_0$: 
$$
\Lambda=\{x+y\mid x\in\Gamma, y\in\Sigma_0\}
$$
The equation $l(x)=c$, for any constant c, defines a 
cross-section of $\Lambda$, which contains exactly 
$A=\vert\Gamma\vert$ vertices. Hence, $\vert \Lambda\vert =(L+1)A$.
We refer to these cross-sections as planes.

\begin{figure}
\begin{center}
\resizebox{!}{7truecm}{\includegraphics{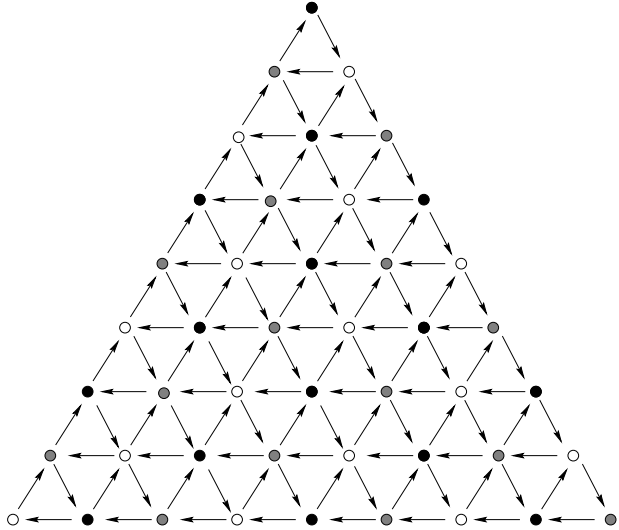}}
\parbox{14truecm}{\caption{\baselineskip=16 pt\small
\label{fig:trilat}
The projection onto the $111$ plane of a cylindrical volume $\Lambda$ with 
triangular base. The shading of the vertices depends on the value of $l(x)$ 
modulo 3. The orientation of the bonds is indicated by arrows. Observe
that each site has an equal number of incoming and outgoing bonds.}
}
\end{center}
\end{figure}

\begin{figure}
\begin{center}
\resizebox{!}{7truecm}{\includegraphics{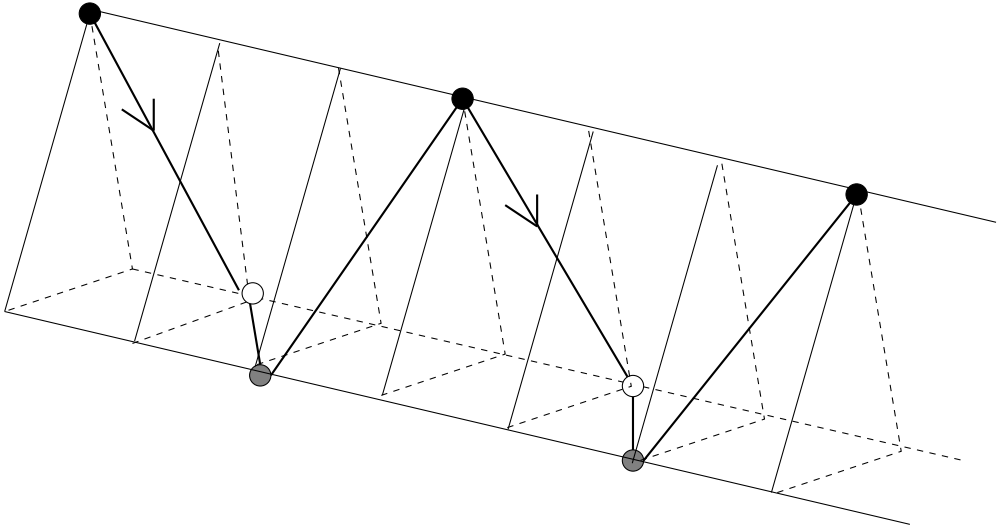}}
\parbox{14truecm}{\caption{\baselineskip=16 pt\small
\label{fig:stick}
The bonds connecting the vertices of a stick $\Sigma$ form a 
one-dimensional subsystem.}
}
\end{center}
\end{figure}

As an example, the projection onto the plane $l(x)=0$, of the vertices of 
$\Lambda$ with triangular base 
is shown in Figure \ref{fig:trilat}, with different shades
depending on the value of $l(x)$ modulo 3.
The orientation of the bonds is indicated by arrows, and one may
observe that each site on the interior of $\Lambda$ 
has an equal number of incoming and outgoing bonds.
By construction, $\Lambda$ can be decomposed into one-dimensional sticks 
running parallel to the cylindrical axis, which we will generically call
$\Sigma$. (See Figure \ref{fig:stick}.) One should observe that $\Sigma$ is
comprised entirely of nearest-neighbor pairs so that every site on $\Sigma$ is
connected to every other site by a sequence of bonds. This will allows us
to exploit the well-known properties of the one-dimensional Heisenberg XXZ 
model to describe $\Sigma$.
The Hamiltonian for the spin-$\frac{1}{2}$ ferromagnetic $XXZ$ Heisenberg
model is given by
\be\label{ham}
H_\Lambda = \sum_{(x_0,x_1)\in B(\Lambda)} h^{q}_{x_0,x_1},
\ee
where 
\be
h^{q}_{x_0,x_1} = - \Delta^{-1} (S_{x_0}^{(1)} S_{x_1}^{(1)}
+ S_{x_0}^{(2)} S_{x_1}^{(2)}) - S_{x_0}^{(3)} S_{x_1}^{(3)} 
+ \frac{1}{4} + \frac{1}{4} A(\Delta) (S_{x_1}^{(3)} - S_{x_0}^{(3)}).
\ee
and $\Delta \geq 1$ is the ``anisotropic coupling'', 
$A(\Delta) = {1 \over 2} \sqrt{1 - 1/\Delta^2}$, and
$q$, $0<q<1$, is the solution of $\Delta = {1 \over 2}(q + q^{-1})$
The matrices $S_x^{(\alpha)}$ ($\alpha = 1,2,3$) are the Pauli spin
matrices acting on the site $x$,
\be
S^{(1)} = \left[\begin{array}{cc} 0 & 1/2 \\
1/2 & 0 \end{array}\right],\quad
S^{(2)} = \left[\begin{array}{cc} 0 & -i/2 \\
i/2 & 0 \end{array}\right],\quad
S^{(3)} = \left[\begin{array}{cc} 1/2 & 0 \\
0 & -1/2 \end{array}\right].		
\ee
The terms containing $A(\Delta)$ cancel on all sites except at the 
top and bottom plane of the cylinder.
The usefulness of the nearest-neighbor Hamiltonian stems from the 
fact that its action on any bond is given by
\begin{eqnarray*}
h^q \dd = 0, && \quad
h^q \du = {1 \over {q + q^{-1}}} 
\left( q \du - \ud \right) ,\\
h^q \uu = 0, && \quad
h^q \ud = - {1 \over {q + q^{-1}}} 
\left( \du - q^{-1} \ud \right).
\end{eqnarray*}
In other words, $h^q$ is the orthogonal projection on the unit vector
\be
\xi_q = {1 \over \sqrt{1 + q^2}} (q \du - \ud). \label{def:xi}
\ee
There is a $(|\Lambda|+1)$-fold degeneracy in the ground states
with a unique ground state for each value of total third component
of the spin $\sum_{x\in\Lambda} S^{(3)}_x$. The basis vectors of the
Hilbert space $(\Cx^2)^{\otimes \vert \Lambda\vert}$ can be labeled with
particle configurations $\alpha =\{\alpha(x)\}_{x \in \Lambda}$, 
where $\alpha(x)$ is 0 or 1, corresponding to $\ket{\uparrow}$ and
$\ket{\downarrow}$, respectively. We write $\Num$ for the operator
defined by
$$
\Num\ket{\alpha} = (\sum_{x \in \Lambda} \alpha(x))\ket{\alpha},
$$
and let $\A(\Lambda,n)$ denote the collection of all 
configurations with $\Num(\alpha) = n$.

Following \cite{ASW} the ground states are given by
\be
\label{def:gs}
\psi_0(\Lambda,n) = \sum_{\alphavec \in \A(\Lambda,n)} 
\bigotimes_{x \in \Lambda} q^{l(x) \alpha(x)} \ket{\alpha(x)} ,
\ee
Note that the weights of $\alpha$ are invariant under 
any permutation of the sites for which planes are invariant. 
These states describe an interface located, on the average, in the
plane determined by $(L/2+l_x)A=n$ \cite{KN2}.

We denote $\Vert \psi_0(\Lambda,n)\Vert^2$ by $Z(\Lambda,n)$.
This quantity is given by
\be\label{uuyy}
Z(\Lambda,n) =\sum_{\alphavec \in \A(\Lambda,n)} \ 
\prod_{x \in \Lambda} q^{2 l(x) \alpha(x)} 
\ee
We will treat $Z(\Lambda,n)$ as a canonical partition function.
It will be useful to consider, also, its grand canonical analogue:
\be
\ZGC{\Lambda}{\mu} = \sum_{n=0}^L Z(\Lambda,n) q^{-2\mu n}
= \prod_{x \in \Lambda} (1 + q^{2 (l(x)-\mu)}) .
\label{grand_canonical}\ee
Then it is easily seen that $\ZGC{\Lambda}{\mu}$ is the squared-norm
of the grand canonical vector defined by
\be\label{def:gc}
\psi^{GC}(\Lambda,\mu) =\sum_{n=0}^{\vert \Lambda\vert}
q^{-n\mu}\psi_0(\Lambda,n)= \bigotimes_{x \in \Lambda}
(\up + q^{l(x)-\mu} \down).
\ee
Due to the product structure, the thermodynamic limit is simply given by
\be
\avgGC{X}{\Ir^3,\mu}=\bigotimes_{x \in \Ir^3} \frac{\bra{\uparrow} 
  + q^{l(x) - \mu} \bra{\downarrow}}{\sqrt{1 + q^{2(l(x)-\mu)}}}
 \;\; X \;\bigotimes_{x \in \Ir^3} \frac{\ket{\uparrow} 
  + q^{l(x) - \mu} \ket{\downarrow}}{\sqrt{1 + q^{2(l(x)-\mu)}}}
\label{product}\ee
for all local observables $X$.

\Section{Equivalence of Ensembles}
\label{sect:eq:ens}

A key step in our argument is the development of an equivalence of ensembles.
Specifically, we will show that for a gauge-invariant local observable
the canonical expectation is close to the grand canonical expectation
for some suitably chosen chemical potential $\mu$.
Here $\mu$ only depends on the total spin of the canonical ensemble, not on
the form of the observable.
From this, naturally follows a thermodynamic limit for gauge-invariant
observables. We begin with activity bounds that show that the ratio of
two canonical partition functions with different particle numbers is
approximately exponential in the difference of the particle numbers,
i.e.,
$$
Z(\Lambda,n-k)\approx Z(\Lambda,n) q^{-2k\mu}
$$
for $\vert k\vert \ll n$. More precisely, we have the following lemma.

{
\lemma[Activity bounds] 
For every volume ${\Lambda}$, 
$|{\Lambda}|=(L+1)A$, the ratio of canonical partition functions
for different number of particles can be bounded from above and 
below by {\rm activity} bounds as follows. Let $A_0$ be any constant.
%Recall the operator $\Num \ket{\alpha} = \ket{\alpha} \sum_x \alpha(x)$,
%and the 1D stick $\Sigma$.
Suppose $n$, $0\leq n \leq A(L+1)$, and $\mu$ are 
such that
\be\label{activ:hyp1}
n - A\navg \leq \frac{1}{2}A_0 A^{1/2}.
\ee
Then, for every $k$ satisfying
\be\label{activ:hyp2}
\vert k\vert\leq \frac{1}{2}A_0 A^{1/2},
\ee
one has the bounds
\be
\frac{Z(\Lambda,n)}{Z(\Lambda,n-k)} \; \le \;
C(A_0,A)
q^{k[2\frac{n}{A}-2\navg+2\mu a\sigma^2-\frac{k}{A}]/(a\sigma^2)},
\label{rrat}
\ee
and
\be
\frac{Z(\Lambda,n)}{Z(\Lambda,n-k)} \; \ge \;
C(A_0,A)^{-1}
q^{k[2\frac{n}{A}-2\navg+2\mu a\sigma^2-\frac{k}{A}]/(a \sigma^2)},
\label{lrat}
\ee
where $a = 2 |\ln q|$,
$$
\sigma^2:=\sigma^2(\mu,L)=\frac{1}{4}\sum_{l=-L/2}^{L/2} 
\frac{1}{\cosh^2(\frac{a}{2}(l-\mu))},
$$
and
\be
C(A_0,A) = \frac{1 + \frac{A_0}{\sigma^2 A^{1/2}}}
{1 - \frac{A_0}{\sigma^2 A^{1/2}}}.
\ee
Moreover, if $\mu$ is the solution of
$ \frac{n}{A}-\navg=0 $,
then, also using the bounds for $\sigma^2$ given in \eq{lower_sigma}, we
obtain
\be
\label{special:rat1}
C(A_0/2,A)^{-1} q^{-\frac{k^2(1-q^2)}{2 a (1+q^2) A}} \leq
q^{-2k\mu} \frac{Z(\Lambda,n)}{Z(\Lambda,n-k)} \leq 
C(A_0/2,A) q^{-\frac{2 k^2(1-q^2)}{a q^2 A}}.
\ee
Alternatively, if $\mu$ solves $\frac{n-k}{A} - \navg = 0$, then we obtain
\be
\label{special:rat2}
C(A_0/2,A)^{-1} q^{\frac{k^2(1-q^2)}{2 a (1+q^2) A}} \leq
q^{-2k\mu} \frac{Z(\Lambda,n)}{Z(\Lambda,n-k)} \leq 
C(A_0/2,A) q^{\frac{2 k^2(1-q^2)}{a q^2 A}}.
\ee
}
\begin{proof}
This can be obtained as follows. Let consider the grand canonical probability 
\be
p(\mu,{\bf n})=q^{-2\mu |{\bf n}|}\frac{Z({\bf n})}{Z_{GC}(\mu)} \, ;
\label{gcprob}
\ee
with
\be
Z({\bf n})=\sum_{\alpha:\A(\Sigma_1,n_1)\otimes\cdots\otimes
\A(\Sigma_{A-A_0},n_{A-A_0})}q^{w(\alpha)}
\label{cprob}
\ee
where $\Sigma_i$ is the i-th one dimensional stick that we are decomposing
our volume in, and where $Z_{GC}(\mu)$ is the grand-canonical partition
function. Clearly, we have
\be
Z(n) \, = \, \sum_{{\bf n}: |{\bf n}|=n} Z({\bf n}) \; .
\ee
Define
\be
p(\mu,n) \, = \, \sum_{{\bf n}: |{\bf n}|=n} p(\mu,{\bf n}) \; ,
\ee
and we have
\be
\frac{Z(n)}{Z(n-k)}=\frac{p(\mu,n)}{p(\mu,n-k)}q^{2k\mu}
\label{ratio}
\ee
The idea now is to make use of the {\it local central limit theorem}
for the probability distribution of the occupation number in the i-th
stick (see \cite{Fel} Theorem XVI.4.3.).
Let $\xi_i=\sum_{x\in\Sigma_i} \alpha_x$. For any integer $N$, consider,  
the probability
\be
P_\mu(\xi_1=n_1,...,\xi_N = n_N) \, = \, p(\mu,{\bf n}) \, .
\label{defpro}
\ee
Due to the factorization property of $p(\mu,{\bf n})$, the $\xi$'s are
independent identically distributed random variables. 
For centered i.i.d. random variables $X_i$ with variance $\sigma^2$,
the local central limit 
theorem guarantees that the random variable
\be
S_N =\frac{1}{\sigma\sqrt{N}}\sum_{n=1}^N X_n\quad.
\label{lclt}
\ee
is close to a Gaussian in the sense that the quantity
\be
P_N(x):=\Prob(\sum_{n=1}^N X_n=x)
\label{cumu}
\ee
fulfills the bounds
\be
\frac{1}{\sigma\sqrt{2\pi N}}e^{-\frac{x^2}{2\sigma^2 N}}
\left(1-\frac{c}{\sqrt{N}}\right)
\leq P_N(x)\leq\frac{1}{\sigma\sqrt{2\pi
N}}e^{-\frac{x^2}{2\sigma^2 N}}\left(1+\frac{c}{\sqrt{N}}\right)
\label{prob_bounds1}
\ee
where $c$ is the constant
\be
c = \frac{\max(|x|,|x-k|)}{\sigma^2 \sqrt{N}} .
\label{ci}
\ee
By applying \eq{prob_bounds1} to the centered quantity 
$X_n=\xi_n - \langle{\xi_n}\rangle$, we obtain the following bounds on the 
ratio of probabilities:
\be
C(N)^{-1}e^{-k(2x-k)/2\sigma^2 N}\leq
\frac{P_N(x)}{P_N(x-k)}\leq C(N)e^{-k(2x-k)/2\sigma^2 N}
\label{ragio}
\ee
where
\be
C(N)=\frac{1+cN^{-1/2}}{1-cN^{-1/2}}\quad .
\label{cicci}
\ee
In terms of the non-centered variables $\xi_i$ we have
\be
p(\mu,n) \; = \; P_A\left(n-A \navg \right)
\label{ncclt}
\ee
where $\navg$ is the average number of particles of a 1D stick $\Sigma$,
in the grand canonical ensemble with chemical potential $\mu$. 	
From this and the hypotheses \eq{activ:hyp1}, \eq{activ:hyp2}, we obtain
\be
c = \frac{A_0}{\sigma^2}
\quad \textrm{ and } \quad
C(A_0,A) = \frac{1 + \frac{A_0}{\sigma^2 A^{1/2}}}
{1 - \frac{A_0}{\sigma^2 A^{1/2}}}.
\ee
Note that in case $\mu$ is chosen so that $\navg = n/A$
or $\navg = (n-k)/A$ then we can replace $c$ by $c/2$,
with the result that $C(A_0,A)$ may be replaced by
$$
C(A_0/2,A) = \frac{1 + \frac{A_0}{2 \sigma^2 A^{1/2}}}
{1 - \frac{A_0}{2 \sigma^2 A^{1/2}}},
$$
as well.
\newline
Also, from \eq{ncclt} and \eq{ragio}, we have
\be
C(A_0,A)^{-1}e^{-\frac{k(2 n - 2 A \navg - k)}{2 \sigma^2 A}}
\leq \frac{p(\mu,n)}{p(\mu,n-k)} 
\leq C(A_0,A) e^{-\frac{k(2 n - 2 A \navg - k)}{2\sigma^2 A}}.
%\label{ragio}
\ee
Using \eq{ratio} (and observing that $q^{2\mu k} = e^{-a \mu}$), 
we have
\be
\frac{Z(n)}{Z(n-k)} \; \le \; 
  C(A_0,A) e^{-k[2\frac{n}{A}-2\navg+2a\sigma^2 \mu - \frac{k}{A}]/2 \sigma^2}\; ,
\label{rratio2}
\ee
and
\be
\frac{Z(n)}{Z(n-k)} \; \ge \; 
  C(A_0,A) e^{-k[2\frac{n}{A}-2\navg+2a\sigma^2 \mu - \frac{k}{A}]/2 \sigma^2}\; .
\label{lratio2}
\ee
Changing to base $q$ then leads to equations \eq{rrat} and \eq{lrat}
of the theorem.
By the derivation of Section \ref{sect:var_n}, we have the bounds
on the variance for the number of particles in a 1D stick:
\be
\label{var:bounds}
\frac{1}{4}\frac{q^2}{1-q^2}\leq \sigma^2(\mu) \leq \frac{1+q^2}{1-q^2}.
\ee
In conjunction with the remark about replacing $C(A_0,A)$ by 
$C(A_0/2,A)$, this gives equations \eq{special:rat1} and \eq{special:rat2}.
\end{proof}

As an application of this lemma, let us consider the case where
$n$ is replaced by $\rho |\Lambda| - n_0$, $k$ is replaced by 
$\rho |\Lambda_0| - n_0$ and $\Lambda$ is replaced by
$\Lambda_0^c := \Lambda \setminus \Lambda_0$.
This means that in the lemma $A$ is replaced by $A - A_0$,
and $(n-k)/A$ is replaced by 
$\rho (|\Lambda| - |\Lambda_0|)/(A-A_0) = \rho (L+1)$.
Then, direct substitution shows
\begin{eqnarray}
&&\frac{Z(\Lambda_0^c,\rho |\Lambda| - n_0)}{Z(\Lambda_0^c,\rho 
|\Lambda_0^c|)} \nonumber\\
&&\qquad \le C(A_0/2,A-A_0)\, q^{-2 k \mu}
e^{-k[2\rho(L+1) - 2 \navg + \frac{k}{A-A_0}]/2 \sigma^2} ,\\
&&\frac{Z(\Lambda_0^c,\rho |\Lambda| - n_0)}{Z(\Lambda_0^c,\rho 
|\Lambda_0^c|)}\; \nonumber\\
&&\qquad \ge C(A_0/2,A-A_0)^{-1}\, q^{-2 k \mu}
e^{-k[2\rho(L+1) -  2 \navg + \frac{k}{A-A_0}]/2 \sigma^2},
\end{eqnarray}
where we have retained $k$, for the moment.
If, further, we choose $\mu$ so that $\navg = \rho (L+1)$, which is
always possible (see Section \ref{sect:avgexp}), then, 
by equation \eq{special:rat2}, we have
\begin{eqnarray}
q^{2 \mu k} 
\frac{Z(\Lambda_0^c,\rho |\Lambda| - n_0)}
{Z(\Lambda_0^c,\rho |\Lambda_0^c|)}\; \le \; 
  C(A_0/2,A-A_0)\, e^{-\frac{k^2}{2 (A-A_0) \sigma^2}} ,\\
q^{2 \mu k} 
\frac{Z(\Lambda_0^c,\rho |\Lambda| - n_0)}
{Z(\Lambda_0^c,\rho |\Lambda_0^c|)}\; \ge \; 
  C(A_0/2,A-A_0)^{-1}\, e^{-\frac{k^2}{2 (A-A_0) \sigma^2}} .
\end{eqnarray}
Using our bounds for $\sigma^2$, we have
\begin{eqnarray}
\label{spef:rat1}
q^{2 \mu k} \frac{Z(\Lambda_0^c,\rho |\Lambda| - n_0)}
{Z(\Lambda_0^c,\rho |\Lambda_0^c|)}\; \le \; 
  C(A_0/2,A-A_0)\, e^{-\frac{(1-q^2) k^2}{2 (1+q^2) (A-A_0)}},\\
\label{spef:rat2}
q^{2 \mu k} \frac{Z(\Lambda_0^c,\rho |\Lambda| - n_0)}
{Z(\Lambda_0^c,\rho |\Lambda_0^c|)}\; \ge \; 
  C(A_0/2,A-A_0)^{-1}\, e^{-\frac{2(1-q^2)k^2}{2 q^2 (A-A_0)}} .
\end{eqnarray}
By our choice of $\mu$, conditions (\ref{activ:hyp1}) and (\ref{activ:hyp2})
are satisfied as long as the order of $L$ does not exceed the order of
$(A-A_0)^{1/2}$.
This estimate will be of use in the next theorem.

Let $\|X\|_{gs}$ denote the operator-norm of $X$ restricted to the 
subspace of ground states. For observables $X$, localized in $\Lambda$
and commuting with $J^{(3)}$, $\|X\|_{gs}$ is also given by
$$
\|X\|_{gs} = \sup_{0\leq n \leq |\Lambda|}
\vert\avg{X}{\Lambda,n}\vert .
$$ 

{\theorem[Equivalence of Ensembles]\label{thm:eqv}
Consider two cylindrical volumes $\Lambda$ and $\Lambda_0$, $\Lambda_0
\subset \Lambda$, of the type defined in Section 2 (in particular
$\vert\Lambda\vert=A (L+1)$, $\vert\Lambda_0\vert=A_0 (L+1)$), and fix a total
number of particles $n_\Lambda$. Define $\rho=n_\Lambda/\vert \Lambda\vert$.
Suppose $X$ is a local observable in the volume $\Lambda_0$, 
which commutes with $J^{(3)} := \sum_x S^{(3)}_x$. Then we have
\be
|\avg{X}{\Lambda,n}
  -  \avgGC{X}{\Lambda_0,\mu}| \; \le \; 
  \varepsilon \|X\|_{gs} \; ,
\label{equicgrc}
\ee
where 
\be
\label{vare}
\varepsilon
  = \frac{\ln^2(A-A_0) + 2 (1+a^2) A_0^2 + 4}{2(A-A_0)} 
  + \frac{4 A_0}{q^2 (A-A_0)^{1/2} - 2 A_0},
\ee
$a = 2 \vert\ln q\vert$,
and the chemical potential $\mu$ is determined by the equation
\be
 \navg = \rho (L+1).
\label{themu}\ee
In particular, for $\rho = 1/2$ the calculations of Section \ref{sect:navg}
will show that $\mu = 0$.
}

{\corollary[Existence of the Thermodynamic limit]\hfill\break
(i) Suppose we have a sequence of pairs $(\Lambda_k,n_k)$ 
with $\Lambda_k$ cylindrical volumes and $\Lambda_k \nearrow \Ir^3$ in such
a way that the length does not grow faster than the linear size of the 
base. Let $\mu_k$ solve $\avgGC{\Num}{\Lambda_k,\mu_k} = n_k$.
Then the convergence $\mu_k \to \mu$ guarantees the convergence,
of $\avg{.}{\Lambda_k,n_k}$ to $\avgGC{.}{\Ir^3,\mu}$, for all local 
observables $X$ commuting with $J^{(3)}$ :
\be \label{thermo:limit}
\avg{X}{\Lambda_k,n_k} \to \avgGC{X}{\Ir^3,\mu} 
\ee
\newline
(ii) Moreover, for any choice of $\mu$, we may find a sequence of pairs
$(\Lambda_k,n_k)$ such that 
\be \label{thermo:limit2}
\avg{X}{\Lambda_k,n_k} \to \avgGC{X}{\Ir^3,\mu} .
\ee
}

\begin{proof}
(Proof of Corollary)
It follows from the monotonicity of $\navg$ proved in Section \ref{sect:navg},
that the equation
\be
\avgGC{\Num}{\Lambda_k,\mu_k} = n_k
\label{mu_eq}\ee
always has a unique solution for $\mu_k$. Then, (i) follows immediately 
from the inequality \eq{equicgrc}, once we observe that $\epsilon \searrow 0$ 
as $\Lambda \nearrow \Ir^3$ in the sense prescribed in the corollary.
\newline
For (ii), take $\Lambda_k$, with base $A_k$, and $n_k$ such that
$$
n_k=\floor{A_k\navg}\quad .
$$
where $\floor{x}$ denotes the largest integer $\leq x$.
Then, $\mu_k$ solving \eq{mu_eq}, is easily seen to converge to $\mu$,
and \eq{thermo:limit2} follows from (i).
\end{proof}

The interpretation of the condition $\mu_k\to\mu$ in (i) of the Corollary
is that, not only does $n_k/\vert\Lambda_k\vert$ converge to $\rho=1/2$,
but, more precisely
$$
n_k=\rho\vert\Lambda_k\vert+\nu A_k + o(A_k)\quad.
$$
The term proportional to $\vert\Lambda_k\vert$ guarantees that the 
interface is in the center of the volume, the second term fixes its
filling factor.

\begin{proof}
(Proof of Theorem \ref{thm:eqv})
Let $\mu$ be determined by \eq{themu}, and define $\Xi$ as follows:
\be
\Xi=\frac{Z(\Lambda, n_\Lambda) q^{-2\mu \rho |\Lambda_0|}}{Z(\Lambda_0^c,
\rho \vert \Lambda_0^c\vert) Z^{GC}(\Lambda_0, \mu)}
\label{Xi}\ee
where $\Lambda_0^c:=\Lambda\setminus\Lambda_0$.
We will obtain the equivalence of ensembles by combining two facts.
The first is that $\Xi$ is approximately equal to $1$, and the second is
an estimate showing that 
$$
\vert \avg{X}{\Lambda,n_\Lambda}\Xi-\avgGC{X}{\Lambda_0,\mu}\vert 
  \leq \varepsilon\Vert X\Vert_{gs} 
$$
But first, let us recall the definitions of the 
expectation of an observable $X$:
\begin{eqnarray}
  \avg{X}{\Lambda,n} 
  &=& \frac{\op{\psi(\Lambda,n)}{X}{\psi(\Lambda,n)}}
  {\ip{\psi(\Lambda,n)}{\psi(\Lambda,n)}}, \\
  \avgGC{X}{\Lambda,\mu} 
  &=& \frac{\op{\psi^{GC}(\Lambda,\mu)}{X}{\psi^{GC}(\Lambda,\mu)}}
  {\ip{\psi^{GC}(\Lambda,\mu)}{\psi^{GC}(\Lambda,\mu)}}.
\end{eqnarray}
Since $X$ is an observable localized in $\Lambda_0$,
we note that $\avgGC{X}{\Lambda,\mu} = \avgGC{X}{\Lambda_0,\mu}$.
Moreover, we may decompose the grand canonical state into
a superposition of canonical states:
\be
\psi^{GC}(\Lambda_0,\mu) 
  = \sum_{n_0 = 0}^{|\Lambda_0|} q^{-\mu n_0} \psi(\Lambda_0,n_0).
\ee
Since $X$ commutes with $J^{(3)}$, it does not have off-diagonal
matrix elements between these canonical states with all different values
of the total spin. Therefore, 
\be
\avgGC{X}{\Lambda,\mu} 
  = \ZGC{\Lambda}{\mu}^{-1} \sum_{n_0 = 0}^{|\Lambda_0|} 
  q^{-2 \mu n_0} Z(\Lambda_0,n_0) \avg{X}{\Lambda_0,n_0}.
\ee
Note also, that since we have a decomposition
\be
\psi(\Lambda,n) = \sum_{n_0 = 0}^{|\Lambda_0|} 
  \psi(\Lambda \setminus \Lambda_0,n-n_0) \otimes 
  \psi(\Lambda_0,n_0),
\ee
and using the previously described properties, we have
\begin{eqnarray}
\avg{X}{\Lambda,n}
  &=& \sum_{n_0 = 0}^{|\Lambda_0|} 
  \frac{Z(\Lambda\setminus\Lambda_0,n-n_0)
  Z(\Lambda_0,n_0)}{Z(\Lambda,n)} \avg{X}{\Lambda_0,n_0} \\
  &=& Z^{GC}(\Lambda_0,\mu)^{-1} \sum_{n_0=0}^{|\Lambda_0|}
  q^{-2\mu n_0} Z(\Lambda_0,n_0)\avg{X}{\Lambda_0,n_0} \times \nonumber\\
  & & \quad \times \frac{Z(\Lambda_0^c,n-n_0) Z^{GC}(\Lambda_0,\mu)}
	{q^{-2\mu n_0} Z(\Lambda,n)}.
\end{eqnarray}
This differs from the definition of $\avgGC{X}{\Lambda_0,\mu}$ only
by the final factor, which is a ratio of partition functions hence amenable
to our activity bounds.
\newline
In fact, we have
\begin{eqnarray}
\avg{X}{\Lambda,n} \Xi - \avgGC{X}{\Lambda,\mu}
  &=& Z^{GC}(\Lambda_0,\mu)^{-1} \sum_{n_0=0}^{|\Lambda_0|}
    q^{-2\mu n_0} \avg{X}{\Lambda_0,n_0} Z(\Lambda_0,n_0) \times \nonumber\\
  & & \quad \times \left[q^{2\mu(n_0 - \avg{n_0}{})} \
    \frac{Z(\Lambda_0^c,n-n_0)}{Z(\Lambda_0^c,\floor{\rho |\Lambda_0|})} 
    - 1\right] 
\end{eqnarray}
where $\avg{n_0}{}=\avgGC{\Num}{\Lambda_0,\mu}$, which equals 
$\rho |\Lambda_0|$ for our choice of $\mu$. Thus we obtain
$|\avg{X}{\Lambda,n} \Xi - \avgGC{X}{\Lambda,\mu}|
  \leq \|X\|_{gs} \avgGC{|g|}{\Lambda_0,\mu},$
where
\be
  g = q^{2\mu(n_0 - \avg{n_0}{})} \
    \frac{Z(\Lambda_0^c,n-n_0)}{Z(\Lambda_0^c,\floor{\rho |\Lambda_0|})} 
    - 1 .
\ee
Now we use the activity bounds \eq{spef:rat1} and \eq{spef:rat2}, 
but replacing $k$ by its actual value, $\avg{n_0}{} - n_0$.
We arrive at the bounds
\begin{eqnarray}
  g &\leq& g_1 := 
  C(A_0/2,A-A_0) e^{-\frac{(1-q^2) (\avg{n_0}{} - n_0)^2}{2 (1+q^2) (A-A_0)}} - 1,\\
  g &\geq& g_2 := 
  C(A_0/2,A-A_0)^{-1} e^{-\frac{2(1-q^2)(\avg{n_0}{} - n_0)^2}{2 q^2 (A-A_0)}} - 1,
\end{eqnarray}
where
\be
C(A_0/2,A-A_0) = \frac{1 + \frac{A_0}{2 \sigma^2 (A-A_0)^{1/2}}}
{1 - \frac{A_0}{2 \sigma^2 (A-A_0)^{1/2}}}.
\ee
Therefore, $|g| \leq \max(|g_1|,|g_2|) \leq |g_1| + |g_2|$.
\newline
We now use the triangle inequality and the fact that the exponent is
negative to obtain:
\be
|g_1| \leq
  \left|1 - e^{-\frac{(1-q^2)(\avg{n_0}{} - n_0)^2}{2 (1+q^2) (A-A_0)}}\right| 
  + |1 - C(A_0/2,A-A_0)|,
\ee
so that
\be
\label{g1:bound1}
\avg{|g_1|}{\Lambda_0,\mu} \leq  
\avgGC{1 - e^{-\frac{(1-q^2) (\avg{n_0}{} - n_0)^2}{2 (1+q^2) (A-A_0)}}}
{\Lambda_0,\mu} + C(A_0/2,A-A_0) - 1.
\ee
Similarly,
\be
\avg{|g_2|}{\Lambda_0,\mu} \leq  
\avgGC{1 - e^{-\frac{2(1-q^2)(\avg{n_0}{} - n_0)^2}{2 q^2 (A-A_0)}}}
{\Lambda_0,\mu} + 1 - C(A_0/2,A-A_0)^{-1}.
\ee
\newline
We will use the Chebyshev inequality to control the expectation term in 
\eq{g1:bound1}. Specifically, for any $B>0$,
\begin{eqnarray*}
\avgGC{1 - e^{-\frac{(1-q^2) (\avg{n_0}{} - n_0)^2}{2 (1+q^2) (A-A_0)}}}
{\Lambda_0,\mu}
	&\leq& \Prob(2 |n_0 - \avg{n_0}{}| \geq 2 B) 
	+ 1 - e^{-\frac{(1-q^2)B^2}{2(1+q^2)(A-A_0)}} \\
	&\leq& q^{2 B} \avgGC{q^{-2 | n_0 - \avg{n_0}{} |}}{\Lambda_0,\mu}
	+ 1 - e^{-\frac{(1-q^2)B^2}{2(1+q^2)(A-A_0)}}.
\end{eqnarray*}
In Section \ref{sect:avgexp} we show that
$\avgGC{q^{-2 |n_0 - \avg{n_0}{}|}}{\Lambda_0,\mu} \leq 2 (2 q^{-2})^{A_0}$.
One choice for $B$ is $a^{-1} [\ln(A-A_0) + A_0 \ln(2 q^{-2})]$.
This gives the bound
\begin{eqnarray}
\avgGC{ 1 - q^{\frac{(n_0 - \avg{n_0}{})^2}{A-A_0}} }{\Lambda_0,\mu}
  &\leq& \frac{2 + \frac{1-q^2}{a^2 (1+q^2)} 
  \left[2 (1+a^2) A_0^2 + \ln^2(A-A_0)\right]}{A-A_0} \nonumber\\
  &\leq& \frac{2 + (1+a^2) A_0^2 + \frac{1}{2} \ln^2(A-A_0)}{A-A_0} \\
  &=:& C_1(A,A_0,q) \nonumber
\end{eqnarray}
The leading order term in the bound is  
$\frac{\ln^2(A-A_0)}{2(A-A_0)}$ for fixed $q$, strictly between 0 and 1.
Also, let 
\be
C_2(q,A,A_0) 
	= \frac{4 A_0}{q^2 (A-A_0)^{1/2} - 2 A_0},
\ee
which is greater than both $C(A_0/2,A-A_0)-1$ and $1 - C(A_0/2,A-A_0)^{-1}$.
Then 
$|\avg{f}{\Lambda,n} \Xi - \avgGC{f}{\Lambda,\mu}|
\leq (C_1 + C_2) \|X\|_{gs}$.
In particular,
$|\avg{\unity}{\Lambda,n} \Xi - \avgGC{\unity}{\Lambda,\mu}|
\leq (C_1 + C_2) \|\unity\|_{gs}$,
which is to say that
$|\Xi - 1| \leq C_1 + C_2$.
Then, using the triangle inequality, we have
\begin{eqnarray*}
|\avg{X}{\Lambda,n} - \avgGC{X}{\Lambda,\mu}|
  &\leq&  |1 - \Xi| \cdot |\avg{X}{\Lambda,n}|
	+ |\avg{X}{\Lambda,n} \Xi - \avgGC{X}{\Lambda,\mu}| \\
  &\leq&  2 (C_1 + C_2) \|X\|_{gs}.
\end{eqnarray*}
So, defining 
$\varepsilon = 2 C_1(q,\Lambda,\Lambda_0,n) + 2 C_2(q,\Lambda,\Lambda_0)$,
the theorem is proved. 
\end{proof}

Note that the restriction to observables $X$ that commute with the
third component of the total spin $J^{(3)}$ is necessary. E.g., the
expectation of $S^+_x$ obviously vanishes in any canonical state,
while it is easy to see, by direct computation, that it does not
vanish in the grand canonical states. This is entirely analogous to
the restriction to gauge invariant observables in particle systems.

%%%%%%%%%%%%%%%%%%%%%%%%%%%%%%%%%%%%%%%%%%%%%%%%%%%%%%%%%%%%%%%%%%%

\Section{Bound on the energy}

In this section we will estimate the energy of a class of perturbations
of the ground state $\psi_0$ given in \eq{def:gs}. Let $\Lambda$ and 
$\Lambda_R$ be two cylindrical volumes as described in Section 2, 
$\Lambda_R\subset\Lambda$. E.g., $\Lambda_R$ and $\Lambda$, may have triangular
cross-sections (see Figure \ref{fig:cylinder}). We will generally assume
that the radius $R$ of $\Lambda_R$ is much less than that of $\Lambda$. 
We consider $\psi$ of the form
\be
\label{ps}
\psi(\Lambda,n,\phi) = \sum_{\alphavec \in \A(\Lambda,n)} 
\bigotimes_{x \in \Lambda} e^{i \phi(x) \alpha(x)} 
q^{l(x) \alpha(x)} \ket{\alpha(x)} ,
\ee
where $\supp(\phi) \subset \Lambda_R$.

We will also suppose that 
\be
\phi = \frac{\Sc}{R} \tilde \phi(\tilde{y}_1,\tilde{y}_2)
\ee
where $\tilde\phi$ is a smooth functions of its variables
and $\Sc$ is a parameter, which we will eventually take to zero 
independent of $R$. The coordinates $\tilde{y}^1,tilde{y}^2$, are
defined by
\be
\tilde{y}^1 = \frac{2 x^1 - x^2 - x^3}{\sqrt{6} R}
\quad {\rm and} \quad \tilde{y}^2 = \frac{x^2 - x^3}{\sqrt{2} R},
\ee
and are  to be viewed as rescaled coordinates for $x$ along the plane 
perpendicular to the 111 axis.

There are two points to our assumptions on $\phi$: First, that $\phi$ is
independent of the 111 component of $x$. Second, that $\phi$ is 
associated to a scale-invariant phase $\tilde\phi$ by 
$\phi(x) = R^{-1} \tilde\phi(x/R)$. Ultimately, the constant $\Sc$ will vanish.
The leading term in our estimate of the gap  is independent of $\Sc$ 
as long as $\Sc \ll 1$. 

Let $\Gamma_R$ be the projection of $\Lambda_R$ onto the plane
$l(x)=0$, $A_R=\vert\Gamma_R\vert$, $\Omega_R$ be the convex hull of 
$\Gamma_R$, and $\tilde{\Omega} = \{x \in\Rl^2: R x \in \Omega_R\}$, 
the rescaled region, and let $m(\tilde{\Omega})$ be the area of 
$\tilde{\Omega}$ (for the standard Lebesgue measure on $\Rl^2$). 

We will also use the following notation: 
$\partial_{\tilde y}\tilde\phi$ and $\partial^2_{\tilde y}\tilde\phi$
are the first- and second-derivative tensors of $\tilde{\phi}$, and by the 
$L^\infty$ norm of a tensor we mean the maximum of the $L^\infty$ norms of 
the components.

Then we have the following theorem.

%%%%%%%% theorem : bound on the energy %%%%%%%%
%%%%%%%% theorem : bound on the energy %%%%%%%%
%%%%%%%% theorem : bound on the energy %%%%%%%%

{\theorem[Bound on $\frac{\op{\psi}{H^{(q)}_{\Lambda}}{\psi}
}{\|\psi\|^2}$] \label{energy:bound}
Considering a perturbed state as in (\ref{ps}), the energy is bounded by
\begin{eqnarray}
\label{bound:en}
\frac{\langle \psi \mid H^{(q)}_{\Lambda} \mid \psi
\rangle}{\|\psi\|^2} \leq
2\frac{1+q^2}{1-q^2}\left(\frac{A_R \Sc^2}{R^4} 
\frac{\|\nabla_{\tilde{y}} \tilde\phi\|_{L^2(\tilde\Omega)}^2}{m(\tilde\Omega)}
+ \mathcal{E}_{\textrm{num}}\right)
\end{eqnarray} 
where
\be
\mathcal{E}_{\textrm{num}} = \frac{6 A_R \Sc^2}{R^5} 
\|\partial_{\tilde y}^2 \tilde\phi\|_{L^\infty}
\|\partial_{\tilde y} \tilde\phi\|_{L^\infty}  
\ee
is a correction to the main term which becomes negligible as $R \to \infty$.
}

\begin{proof}
We begin by calculating how a two-site hamiltonian $h_b^q$ 
acts on the perturbed state. 

We consider the decomposition of our lattice into the relevant bond
$b = (x_0,x_1)$ and everything else $\Lambda \setminus b$.
Thus
\be
  h_b^q = \unity_{\Lambda \setminus b} \otimes \ket{\xi_b} \bra{\xi_b},
\ee
where $\xi_b$ is the unit vector from (\ref{def:xi}) on the 
pair $b$, and
\be
\psi(\Lambda,n)=\sum_{n_b=0}^2 \psi(\Lambda \setminus b,n-n_b) \otimes
\psi(b,n_b).
\ee
Here $\psi(b,n_b)$ is as would be defined by (\ref{ps}),
but with $\Lambda$ replaced by $b$ and $n$ replaced by $n_b$.
For example $\psi(b,1) = q^{l(x_0)} e^{i \phi(x_0)} \du 
+ q^{l(x_1)} e^{i \phi(x_1)} \ud$. But $\xi_b$ is orthogonal to 
$\psi(b,0)$ and $\psi(b,1)$, since $\xi_b$ lies in the sector of total spin 1.
And
\be
\ip{\xi_b}{\psi(b,1)} = \frac{1}{\sqrt{1+q^2}} 
q^{l(x_0)+1} e^{i \phi(x_0)} (1 - e^{i [\phi(x_1) - \phi(x_0)]}).
\ee
Now it is straightforward to see
\begin{eqnarray}
&&\op{\psi(\Lambda,n)}{h_b^q}{\psi(\Lambda,n)}\\
&&\quad =\|\psi(\Lambda \setminus b,n-1)\|^2\, |\ip{\xi_b}{\psi(b,1)}|^2 
\nonumber \\
&&\quad =\frac{2}{(q+q^{-1})^2}Z(\Lambda,n)P^{q}(b) (1 - \cos[\phi(x_1) - \phi(x_0)]),
\end{eqnarray}
where we have defined 
\be
P^q(b) = \frac{Z(\Lambda \setminus b,n-1)Z(b,1)}{Z(\Lambda,n)}.
\ee
Then we may write
\be\label{ene}
\frac{\langle \psi \mid H^{(q)}_{\Lambda} \mid \psi \rangle}{Z(\Lambda,n)}
= \frac{2}{(q+q^{-1})^2}~ \sum_{b \in B(\Lambda)} \ P^q(b)
(1 - \cos[\phi(x_1) - \phi(x_0)]).
\ee
Actually, $P^q(b)$ depends on $b$ only through $l(x_0)$.
So from here on, we'll write it as $P^q(l(x_0))$, and observe the following:
\be\label{pb}
\frac{\langle \psi \mid H^{(q)}_{\Lambda} \mid \psi \rangle}{Z(\Lambda,n)}
= \frac{2}{(q+q^{-1})^2} \sum_{l=-L/2}^{L/2-1} P^q(l) 
  \sum_{x \in \Gamma^l_R} \sum_{j=1}^3 
  (1 - \cos[\phi(x+e_j) - \phi(x)]) ,
\ee
where $\Gamma^l_R = \{x \in \Lambda_R : l(x) = l\}$.

Let us estimate the term
$\sum_{x \in \Gamma^l_R} \sum_{j=1}^3 (1 - \cos[\phi(x+e_j) - \phi(x)])$.
We have an inequality
\be
\label{cos:ineq}
 1 - \cos[\phi(x+e_j) - \phi(x)] \leq \frac{1}{2} [\phi(x+e_j) - \phi(x)]^2
\ee
(which is actually an  equality in the limit $R \to \infty$ for our ansatz).
Also,
\be
\label{diff:approx}
\sum_{i=1}^3 [\phi(x+e_j) - \phi(x)]^2 \approx |\nabla_x \phi(x)|^2 
	= \frac{\Sc^2}{R^4} |\nabla_{\tilde y} \tilde \phi|^2
\ee
In fact, using the inequality
\be
|[\tilde\phi(\tilde{y}+v) - \tilde\phi(\tilde y)]^2 - 
[c\cdot\nabla_{\tilde y} \tilde\phi(\tilde{y})]^2|
  \leq \| \partial_{\tilde y}^2 \tilde\phi \|_{L^\infty}
  \| \partial_{\tilde y} \tilde\phi\|_{L^\infty}
  \|v\|_{l^1}^3
\ee
one may conclude that the error in (\ref{diff:approx}) is bounded by 
$\frac{3 \Sc^2}{R^5} \|\partial_{\tilde y}^2 \tilde\phi\|_{L^\infty}
\|\partial_{\tilde y} \tilde\phi\|_{L^\infty}$.

Incorporating this estimate into the inequality of (\ref{cos:ineq}),
we have
\begin{eqnarray}
&& \sum_{x \in \Gamma^l_R} \sum_{j=1}^3 (1 - \cos[\phi(x+e_j) - \phi(x)]) 
  \leq \qquad \qquad \nonumber\\
&& \qquad \qquad 
\frac{1}{2 R^2} \sum_{x \in \Gamma^l_R} 
  |\nabla_{\tilde{y}} \phi(x)|^2
  + \frac{3 \Sc^2 |\Gamma^l_R|}{2 R^5} 
  \|\partial_{\tilde y}^2 \tilde\phi\|_{L^\infty}
  \|\partial_{\tilde y} \tilde\phi\|_{L^\infty} 
  \label{Energ:Ineq1} 
\end{eqnarray}
Finally, as $R \to \infty$, the sum over each $\Gamma^l_R$ becomes increasingly
well-approximated by the integral over $\Omega_R$,
we is proved in Lemma \ref{lem:voronoi} immediately following this proof.
The lemma gives us a bound
\be
\sum_{x \in \Gamma^l_R} 
|\nabla_{\tilde{y}} \phi(x)|^2 
  \leq \frac{\Sc^2 |\Gamma^l_R|}{R^2} \left[\frac{1}{m(\tilde{\Omega})} 
  \int_{\tilde{\Omega}} |\nabla_{\tilde{y}} \tilde\phi|^2\, d^2 y 
  + \frac{\rho}{R} 
  \|\nabla^2_{\tilde{y}} \tilde\phi \nabla_{\tilde{y}} \tilde\phi
	\|_{L^\infty(\tilde{\Omega})}\right] ,
  \label{Energ:Ineq2}
\ee
where $\nabla^2$ is the Laplacian and $\rho = \sqrt{2/3}$ is the
maximum radius for the Voronoi domain.
(Note that by its definition, as the trace of the second-derivative tensor,
the Laplacian enjoys the bounds
\be
\|\nabla^2_{\tilde{y}} \tilde\phi \nabla_{\tilde{y}} \tilde\phi
  \|_{L^\infty(\tilde{\Omega})} 
  \leq 2 \|\partial_{\tilde y}^2 \tilde\phi\|_{L^\infty}
  \|\partial_{\tilde y} \tilde\phi\|_{L^\infty}, 
\label{lasteq}\ee  
which may be combined with error term in (\ref{Energ:Ineq1}).)
Combining \eq{Energ:Ineq2} and \eq{lasteq} gives us the theorem, modulo
the term $\sum_{l=-L/2}^{L/2-1} P^q(l)$, for which we derive the necessary
in Lemma \ref{lem:Pq}.
\end{proof}

\begin{lemma}\label{lem:voronoi}
Suppose $\Gamma$ is a region in a regular lattice.
For each $x \in \Gamma$, let $\Omega_x$ be the Voronoi domain of $x$ with
respect to the whole lattice, and
let $\Omega_\Gamma$ be the union of all the individual domains $\Omega_x$.
If $f$ is a smooth function on $\Omega_\Gamma$, then
\be
\left| \frac{1}{|\Gamma|} \sum_{x \in \Gamma} f(x)
	- \frac{1}{m(\Omega_\Gamma)} \int_{\Omega_\Gamma} f(y)\, dy \right|
	\leq \rho \|\nabla_y f\|_{L^\infty(\Omega_\Gamma)}
\ee
where $\rho$ is the maximum radius of a Voronoi domain.
\end{lemma}

\begin{proof}
For each $x \in \Gamma$,
\begin{eqnarray*}
f(x) - \frac{1}{m(\Omega_x)} \int_{\Omega_x} f(y)\, dy 
  &\leq& -\frac{1}{m(\Omega_x)} \int_{\Omega_x} [f(y) - f(x)]\, dy \\
  &=& -\frac{1}{m(\Omega_x)} \int_{\Omega_x} \int_0^1 
    \frac{d}{dt} f(x + t (y-x))\, dt\, dy \\
  &=& -\frac{1}{m(\Omega_x)} \int_{\Omega_x} \int_0^1 
    \nabla_y f(x + t(y-x)) \cdot (y-x)\, dt\, dy .
\end{eqnarray*}
This clearly leads to the bound
\be
\left| f(x) - \frac{1}{m(\Omega_x)} \int_{\Omega_x} f(y)\, dy \right|
	\leq \rho(\Omega_x) \|\nabla_y f\|_{L^\infty(\Omega_x)}.
\ee
From this, the lemma follows easily.
\end{proof}

Now, we will derive the necessary bound on
$$
\sum_{l=-L/2}^{L/2-1} P^q(l)\quad .
$$
We will rely on bounds for similar quantities in the 
one-dimensional model proved in \cite{BCN}.

\begin{lemma}[Bound on $\sum_{l=-L/2}^{L/2-1} P^q(l)$]\label{lem:Pq}
\be
\sum_{l=-L/2}^{L/2-1} P^q(l) \leq  2\frac{1+q^2}{1 - q^2} .
\ee
\end{lemma}

\begin{proof}
Recall
\be
P^q(l)=\frac{Z(\Lambda \setminus b,n-1) Z(b,1)}
{Z(\Lambda,n)}.
\ee
The ratio of partition functions in the equation above is clear:
It is the probability of finding one particle shared by the 
sites of $b$, and $n-1$ particles shared by the sites of 
$\Lambda \setminus b$, conditioned on finding $n$ total particles on 
$\Lambda$. We consider the operator
$$Y_b = \unity_{\Lambda \setminus b} \otimes 
  \left(\ket{\uparrow \downarrow}_b \bra{\uparrow \downarrow}_b
  + \ket{\downarrow \uparrow}_b \bra{\downarrow \uparrow}_b\right).$$
Then
\be
\frac{Z(\Lambda \setminus b,n-1) Z(b,1)}{Z(\Lambda,n)}
	= \avg{Y_b}{\Lambda,n},
\ee
and
\be
\sum_{l=-L/2}^{L/2-1} P^q(l)
  =\left\langle\sum_{l=-L/2}^{L/2-1} Y_{b(l)}
  \right\rangle_{\Lambda,n}.
\label{PtoY}\ee
where $b(l)= (x_0,x_1)$, where $l(x_0)=l$, and $(x_0,x_1)$ is a bond in the
stick containing the origin, which we denote by $\Sigma_0$.
The restriction of the state in $\Lambda$ with $n$ spins down is of the
form
$$
\langle X\rangle_{\Sigma_0}=\sum_{k=0}^{L+1} c_k \avg{X}{\Sigma_0,k}
$$
where $X$ is any observable commuting with $J^{(3)}=\sum_{x\in\Sigma_0}
S^{(3)}_x$, as is, e.g., $Y_{b(l)}$, and the $c_k$ are non-negative
numbers summing up to one. We will now derive an upper bound for 
$\avg{\sum_{l=-L/2}^{L/2-1} Y_l}{\Sigma_0}$, that is independent of the
coefficients $c_k$.
We start from
\be
\avg{Y_l}{\Sigma_0,k}
\leq \Prob_k( S^{(3)}_l=\uparrow, S^{(3)}_{l+1}=\downarrow)
+\Prob_k( S^{(3)}_l=\downarrow, S^{(3)}_{l+1}=\uparrow)
\label{oneDprob}\ee
where $\Prob_k$ denotes the probability in the ground state with $k$
spins down for a one-dimensional system on $[-L/2,L/2]$, the sites of which
we label by $l$. Each term in the RHS of \eq{oneDprob} can be estimate as
follows.
\be
\Prob_k( S^{(3)}_l=\uparrow, S^{(3)}_{l+1}=\downarrow)
\leq
\min\left(\Prob_k(S^{(3)}_l=\uparrow),\Prob_k(S^{(3)}_{l+1}=\downarrow)\right)
\ee
Theorem 7.1 of \cite{BCN} gives the following bounds
\beann
\Prob_k(S^{(3)}_{l+1}=\downarrow)\leq q^{2(l-(k+1-L/2)}
&\mbox{if}& l\geq k+1-L/2\\
\Prob_k(S^{(3)}_l=\uparrow)\leq q^{2(k+1-L/2-l)}
&\mbox{if}& l < k+1-L/2
\eeann
Combining these inequalities and summing over $l$ yields
\be
\sum_{l=-L/2}^{L/2-1} \avg{Y_l}{\Sigma_0,k}\leq 2\frac{1+q^2}{1-q^2}
\ee
for all $k=0,\ldots,L+1$. Together with \eq{PtoY} this concludes the proof.
\end{proof}

%%%%%%%%\Section{Bound for the denominator}%%%%%%%%
%%%%%%%%\Section{Bound for the denominator}%%%%%%%%
%%%%%%%%\Section{Bound for the denominator}%%%%%%%%
\Section{Bound for the denominator}

Note that $\psi(\Lambda,n)=T(\phi)\psi_0(\Lambda,n)$, where $T(\phi)$ is 
the unitary operator defined by,
\be
T(\phi) = \bigotimes_{x \in \Lambda} (\ket{\uparrow}\bra{\uparrow}
  + e^{i \phi(x)} \ket{\downarrow} \bra{\downarrow}).
\ee
In particular, $\Vert T(\phi)\psi_0(\Lambda,n)\Vert^2=
\Vert\psi(\Lambda,n)\Vert^2=Z(\Lambda,n)$. For convenience, we will
sometimes omit the arguments $\Lambda$ and $n$ from the notation.
In this section we will consider the half-filled system, i.e,
$\rho=n/\vert\Lambda\vert=1/2$. This corresponds to $\mu=0$. 

\begin{theorem}[Bound on $|\frac{<\psi_0|\psi>}{<\psi_0|\psi_0>}|$ ]
\label{den:bounds}
Considering a perturbed state in the volume $\Lambda_0$ defined by
(\ref{ps}) we have that canonical and grand-canonical expectations
of the perturbed state are arbitrarily close for large volumes $\Lambda$
in the sense:
\be
\label{gcbl}
\left\vert\frac{\ip{\psi}{\psi_0}}{\ip{\psi_0}{\psi_0}}
  - \avgGC{T(\phi)}{\Lambda,\mu} \right\vert
  \le \frac{\ln^2(A-A_0) + 2 (1+a^2) A_0^2 + 4}{2(A-A_0)} 
  + \frac{4 A_0}{q^2 A^{1/2} - 2 A_0}.
\ee
Moreover, with the ansatz defined by (\ref{ps}), the grand canonical 
expectation is bounded as
\begin{eqnarray}
\label{deb}
&&\ln \left|\avgGC{T(\phi)}{\Lambda,\mu}\right|^2
  \leq \qquad \qquad \\
&& \qquad \qquad \leq - q^{2 \delta(\mu)} \frac{A_R \Sc^2}{4R^2} 
  \Bigg[\frac{\|\tilde\phi\|_{L^2(\tilde\Omega)}^2}{m(\tilde\Omega)} 
  - \frac{\sqrt{6}}{R} \|\partial_{\tilde{y}} \tilde\phi \|_{L^\infty} 
    \|\tilde\phi\|_{L^\infty}
  - \frac{\Sc^2}{12 R^2} \|\tilde\phi\|_{L^\infty}^4 
\Bigg] \nonumber
\end{eqnarray}
where $\delta(\mu)$ is the distance of $\mu$ from its closest integer
neighbor.
(Recall that we have defined the $L^\infty$-norm of a tensor 
to be the $L^\infty$-norm of its maximum component.)
\end{theorem}

\begin{proof}
The proof of equation (\ref{gcbl}) is a direct consequence of the 
equivalence of ensembles because, since $T(\phi)$ is a unitary operator,
$\|T(\phi)\| = 1$. Let us now consider the proof of equation (\ref{deb}).

We wish to bound the denominator from below;
i.e.\ to demonstrate that $1 - |\avg{T(\phi)}{\Lambda,n}|^2$ is not too small.
This is tantamount to showing that $|\avg{T(\phi)}{\Lambda,n}|^2$
is not too close to 1.
Furthermore, we know this quantity 
lies between 0 and 1.
We estimate the actual canonical average with the grand canonical average,
and take the logarithm in order to exploit the factorization properties
of the grand canonical ensemble.
First, we note
\be\label{rat}
\left|\avgGC{T(\phi)}{\Lambda,\mu}\right|
  = \left| \prod_{x \in \Lambda_0} \frac{1+e^{i \phi(x)} q^{2 (l(x)-\mu)}} 
  {1+q^{2 (l(x)-\mu)}} \right|.
\ee
Recall the definition $a=-2 \ln{q}$. 
This allows us a more convenient form
in place of (\ref{rat})
\begin{eqnarray}
\label{uu}
&&\left| \prod_{x \in \Lambda_0} \frac{1+e^{i \phi(x)} q^{2 (l(x)-\mu)}}
  {1+q^{2 (l(x)-\mu)}} \right|^2 \nonumber\\
  &&\quad = \prod_{x \in \Lambda_0} 
  \frac{e^{2 a (l(x)-\mu)} + 2 \cos \phi(x) e^{a (l(x)-\mu)} + 1}
  {e^{2 a (l(x)-\mu)} + 2 e^{a (l(x)-\mu)} + 1} \\
  &&\quad = \prod_{x \in \Lambda_0} 
  \left(1 - \frac{1}{2}(1-\tanh^2[a (l(x)-\mu)/2])
	(1 - \cos \phi(x))\right).
\nonumber
\end{eqnarray}
We partition the product over planes and estimate the logarithm, thus:
\begin{eqnarray*}
\ln \left|\avgGC{T(\phi)}{\Lambda,\mu}\right|^2
  &=& \ln \left( \prod_{x\in \Lambda_0} 1 - \frac{1}{2}
  (1-\tanh^{2}[a (l(x)-\mu)/2])(1 - \cos \phi(x)) \right ) \\
  &\leq& - \frac{1}{2} \sum_{x \in \Lambda-0} (1 - \tanh^2[a (l(x)-\mu)/2]) 
	(1 - \cos \phi(x)) \\
  &=& - \frac{1}{2} \sum_{l=-L/2}^{L/2} (1 - \tanh^2[a (l-\mu)/2])
	\sum_{x \in \Gamma^l_R} (1 - \cos \phi(x)) .
\end{eqnarray*}
We may approximate $1 - \cos(\phi(x))$
by $\frac{1}{2} \phi(x)^2$, with an error no larger than  
$\frac{1}{24} \|\phi\|_{L^\infty}^4 $
which is the same as $ \frac{\Sc^4}{24 R^4} \|\tilde\phi\|_{L^\infty}^4$.
In this case
\be
\label{sumsum}
\ln \left|\frac{Z_{GC}(\Lambda_0,\mu,\phi)}{Z_{GC}(\Lambda_0,\mu,0)}\right|^2
\leq - \frac{1}{2} \sum_{l=-L/2}^{L/2} (1 - \tanh^2 [a (l-\mu)/2])
  \left[\sum_{x \in \Gamma^l_R} \frac{1}{2} \phi_x^2 
  - \frac{\Sc^4 |\Gamma^l_R|}{24 R^4} \|\tilde\phi\|_\infty^4 \right] .
\ee
We may approximate the sum over $\Gamma^l_R$ with an integral such that 
the error is bounded by 
$\frac{\rho \Sc^2 |\Gamma^l_R|}{R^3} 
\|\nabla_{\tilde{y}} \phi \|_{L^\infty} \|\tilde\phi\|_{L^\infty}$.
We may bound the sum 
$\sum_{l=-L/2}^{L/2} (1 - \tanh^2[a (l-\mu)/2])$
from below by its largest term (since all the terms are positive).
The largest term occurs for that integer $l$ which is closest to $\mu$.
Thus, defining 
$\delta(\mu) = \min(\mu - \lfloor{\mu}\rfloor,\lceil{\mu}\rceil - \mu)$, 
we see
\be
\sum_{l=-L/2}^{L/2} (1 - \tanh^2[a(l-\mu)/2])  
  \geq 1 - \tanh^2[a \delta(\mu)/2] 
  = \frac{4}{(q^{\delta(\mu)} + q^{-\delta(\mu)})^2} 
  \geq q^{2\delta(\mu)},
\ee
Using these bounds, we may continue the estimate of $\eq{sumsum}$.
We arrive at
\begin{eqnarray}
\label{den:exp:bound}
&&\ln \left|\avgGC{T(\phi)}{\Lambda,\mu}\right|^2
  \leq \qquad \qquad \\
&& \qquad \qquad \leq - q^{2 \delta(\mu)}
  \frac{\Sc^2 |\Gamma^l_R| }{4R^2} 
  \Bigg[\frac{\|\tilde\phi\|_{L^2(\tilde\Omega)}^2}{m(\tilde\Omega)} 
  - \frac{\rho}{R} \|\nabla_{\tilde{y}} \tilde\phi \|_{L^\infty} 
    \|\tilde\phi\|_{L^\infty}
  - \frac{\Sc^2}{12 R^2} \|\tilde\phi\|_{L^\infty}^4 
\Bigg] \nonumber
\end{eqnarray}
Since $|\nabla_{\tilde y}\tilde\phi| 
\leq 2 \|\partial_{\tilde y}\tilde\phi\|_{l^\infty}$
and since $\rho = \sqrt{3/2}$, we have equation \eq{deb}.
\end{proof}

%%%%%%%%\subsection{Bound on the Ratio}%%%%%%%%
%%%%%%%%\subsection{Bound on the Ratio}%%%%%%%%
%%%%%%%%\subsection{Bound on the Ratio}%%%%%%%%
\subsection{Bound on the Ratio}
We will now combine the results of the bound on the numerator and the bound 
on the denominator to get a true bound on the spectral gap.
We first allow $\Lambda \nearrow \Ir^3$ in the appropriate fashion so that
$\varepsilon \searrow 0$.
Then we consider the case that $S\to 0$, holding $R$ fixed.
This means that we consider a perturbation to the ground state which is 
very small.
But since the ground state has energy zero, the energy of the 
perturbed state is entirely due to the small perturbation.
In fact it is proportional to the size of the perturbation, and from this we 
obtain a linearized (with respect to amplitude of $\phi$) bound: 
In fact we have, combining \eq{vp}, \eq{bound:en}, and \eq{gcbl}
\be
\label{rb}
\gamma_1 \leq \frac{16 q^{2(1-\delta(\mu))}}{(1-q^2)R^2} \cdot
  \frac{\|\nabla_{\tilde{y}}\tilde\phi\|_{L^2(\tilde\Omega)}^2/m(\tilde\Omega) 
  + \frac{6}{R} \|\partial_{\tilde y}^2\tilde\phi\|_\infty 
  \|\partial_{\tilde y}\tilde\phi\|_\infty}
  {\|\tilde\phi\|_{L^2(\tilde\Omega)}^2/m(\tilde\Omega) - \frac{\sqrt{6}}{R} 
  \|\partial_{\tilde y}\tilde\phi\|_\infty
  \|\tilde\phi\|_\infty}
\ee
Note that this bound is homogeneous with respect to the amplitude of $\phi$,
which is the result of our linearization.
We observe that, whatever the form for $\tilde\phi$, as long as it is smooth
we have the same asymptotic behavior for the bound on the spectral gap.
Namely $\gamma_1 = O(1/R^2)$.
This said, it is certainly worthwhile to find a best bound, which we 
take up presently.

%%%%%%%%\subsection{The Bessel Function Ansatz}%%%%%%%%
%%%%%%%%\subsection{The Bessel Function Ansatz}%%%%%%%%
%%%%%%%%\subsection{The Bessel Function Ansatz}%%%%%%%%

\subsection{The Bessel Function Ansatz}
Let us write the leading-order term in the bound for the spectral gap:
\be
E(\tilde\phi) 
  = \frac{\|\nabla_{\tilde{y}}\tilde\phi\|_2^2}{\|\tilde\phi\|_2^2}.
\ee
In order to minimize the bound on the spectral gap, we will minimize
the functional $E(\phi)$ amongst all functions $\phi$ which possess two
continuous derivatives and which vanish on the boundary of 
the rescaled perturbed region $\tilde\Omega$.
(In order that the ``small'' phase $\phi$ match the external phase 
of $0,\pm2 \pi,\dots$ on $\partial\Omega$, it must be zero there.
Thus $\tilde\phi \equiv 0$ on $\partial\tilde\Omega$.)
Therefore, we consider the first variation
\be
\lim_{\tau \to 0} \frac{1}{\tau}[E(\phi+\tau \phi') - E(\phi)] 
  = \frac{2 \int \nabla\phi \cdot \nabla\phi'}{\int \phi^2}
  - \frac{2 \int \phi \phi' \int |\nabla \phi|^2}{\int\phi^2 \int\phi^2}.
\ee
Setting the first variation to zero for all test functions 
$\phi'$ leads to the eigenvalue problem for Laplace's equation
\be
\label{pde:1}
\left\{ \begin{array}{ll}
  -\nabla^2 \tilde\phi = \lambda \tilde\phi 
  & \textrm{ in } \tilde\Omega, \\
  \tilde\phi = 0 & \textrm{ on } \partial\tilde\Omega,
  \end{array} \right.
\ee
where $\lambda = E(\phi)$.

We choose, for our domain, the unit disk.
We seek the solution to equation (\ref{pde:1}) which minimizes $\lambda$,
but with the restriction that $\phi$ must possess two continuous derivatives.
So the fundamental solution, which is the logarithm,  is disallowed
(and, in fact, has higher energy).
We seek the first eigenstate of the Laplacian above the ground state.
This is a classic problem, found in any elementary PDE text, 
with the Bessel Function for the solution:
$$\tilde\phi(\tilde y) = J_0(z_0 r) ,$$
where $r = |\tilde y|$, $J_0$ is the zeroth Bessel function,
and $z_0 \approx 2.406$ is its first zero.
Now, using this choice for $\phi$ and the bounds \eq{rb}, we obtain
\be
\gamma_1 \leq \frac{16 q^{2(1-\delta(\mu))}}{(1-q^2)R^2} \cdot
  \frac{1.56 + \frac{6}{R} (2.90)(1.40)}
  {0.27 - \frac{\sqrt{6}}{R} (1.40)(1)}.
\ee
Thus,
\be
\gamma_1 \leq \frac{100 q^{2(1-\delta(\mu))}}{(1-q^2)R^2} 
\quad \textrm{for}
\quad R>70.
\ee

\Section{Results from the 1D grand canonical ensemble}

\subsection{The mean number of particles in a stick}
\label{sect:navg}
Recall that $\Sigma$ is a 1D stick running parallel to the 111 axis.
So, it is actually a 1D spin chain.
We wish to estimate the mean number of particles in
$\Sigma$, for the grand canonical ensemble.
This is
\begin{eqnarray}
  \navg := \ZGC{\Sigma}{\mu}^{-1}
	\sum_{n=1}^{L+1} n q^{-2 \mu n} Z(\Sigma,n)\\ 
	= \ZGC{\Sigma}{\mu}^{-1}
	\sum_{n=1}^{L+1} n e^{a \mu n} Z(\Sigma,n) .\nonumber
\end{eqnarray}
where $\Sigma$ is the interval 
$\{-\frac{L}{2},-\frac{L}{2}+1,\dots,\frac{L}{2}\}$.
(Recall $a = - 2 \log q$.)
By a standard calculation, we have
\be
\navg = \frac{1}{a} \frac{\partial}{{\partial \mu}} 
\log \ZGC{\Sigma}{\mu} .
\ee
On the other hand, the grand canonical partition function factorizes, 
as we have seen, so that
\be 
  \navg = \sum_{l=-L/2}^{L/2} 
	{e^{a(\mu - l)} \over {1 + e^{a(\mu - l)}}}
	= \sum_{l=-L/2}^{L/2} \frac{1}{2} 
	\left[1 - \tanh\left(\frac{a}{2} (l - \mu)\right)\right].
\ee
An examination of the graph of the function $x \mapsto 1 - \tanh(x)$
reveals an approximate heaviside function, with support on the negative axis.
We define the function 
\be
\eta(x) = \left\{ 
  \begin{array}{ll}
  1 & x < 0,\\
  1/2 & x = 0,\\
  0 & x > 0.\end{array} \right. 
\ee
Then, as long as $-L/2 \leq \mu \leq L/2$, we remark
\be
\navg = 
  \left\{ \begin{array}{ll}
  \floor{\mu} + \frac{L}{2} & \mu \not\in \Ir,\\
  \mu + \frac{L+1}{2} & \mu \in \Ir\end{array} \right\} 
  +\sum_{l=-L/2}^{L/2} \left({1 \over 2} 
  - {1 \over 2} \tanh\left({a \over 2}(l - \mu)\right) - \eta(l - \mu)\right).
\ee
We make the definition
\be
F_L(\mu) = \navg - \left(\mu + \frac{L+1}{2}\right) 
\ee
For $\mu$ in the range above one may determine (by combining the two tails
in the series and estimating upwards by an integral) that
\be 
|F_\infty(\mu) - F_L(\mu)| 
  \leq \frac{1}{a} \ln\left(\frac{1 + \exp(-\frac{a}{2}(\frac{L}{2} - \mu))}
  {1 + \exp(-\frac{a}{2}(\frac{L}{2} + \mu))}\right)
\ee
Notice that in case $\mu = 0$, there is no error at all in estimating 
$F_L$ by $F_\infty$,
and, furthermore, $F_\infty(0)=0$.
It is clear that $F_\infty(\mu)$ is periodic in $\mu$ with period 1, 
because it is a sum over the entire integer lattice, so it will suffice for 
us to consider $\mu$ in the range $]0,1[$.
A straightforward calculation then yields
\begin{eqnarray*}
F_\infty(\mu) 
  & = & -\mu + \frac{1}{2} - \frac{1}{1 + e^{a \mu}}
  + \sum_{l=1}^\infty \left[ \frac{1}{1 + e^{a(l - \mu)}} -
  \frac{1}{1 + e^{a(l + \mu)}} \right] \\
  & = & -\mu + \frac{1}{2}\tanh(a \mu)
  + \sum_{l=1}^\infty \frac{\sinh(a \mu)}
	{\cosh(a \mu) + \cosh(a l)}  
\end{eqnarray*}
Defining $\{\mu\} = \mu - \floor{\mu}$ we have
\be
F_\infty(\mu) = - \{\mu\} + \frac{1}{2}\tanh(a \{\mu\})
  + \sum_{l=1}^\infty \frac{\sinh(a \{\mu\})}
	{\cosh(a \{\mu\}) + \cosh(a l)}
\label{defF}\ee
for all values of $\mu$.

\begin{lemma}
\label{lem:avgn}
The function $F_\infty$ defined in \eq{defF} has the following properties:
i) $F_\infty$ is periodic with period $1$, i.e, 
$F_\infty(\mu+1)=F_\infty(\mu)$, for all
$\mu\in\Rl$.\newline
ii) $F_\infty$ is odd about $\mu=1/2$, i.e., 
$F_\infty(1-\mu)=-F_\infty(\mu)$, for
all $\mu\in\Rl$.\newline
iii) $-1\leq F_\infty(\mu)\leq 1$, for all $\mu\in\Rl$.\newline
iv) $F_\infty(\mu) = 0$ for $\mu \in \Z$ and $\mu \in \frac{1}{2} + \Z$.
I.e.\ the estimate $\navg = \mu + \frac{L+1}{2}$ is exact for 
half-integer and integer filling.
\end{lemma}

\begin{figure}[t]
\begin{center}
\resizebox{13cm}{!}{\includegraphics{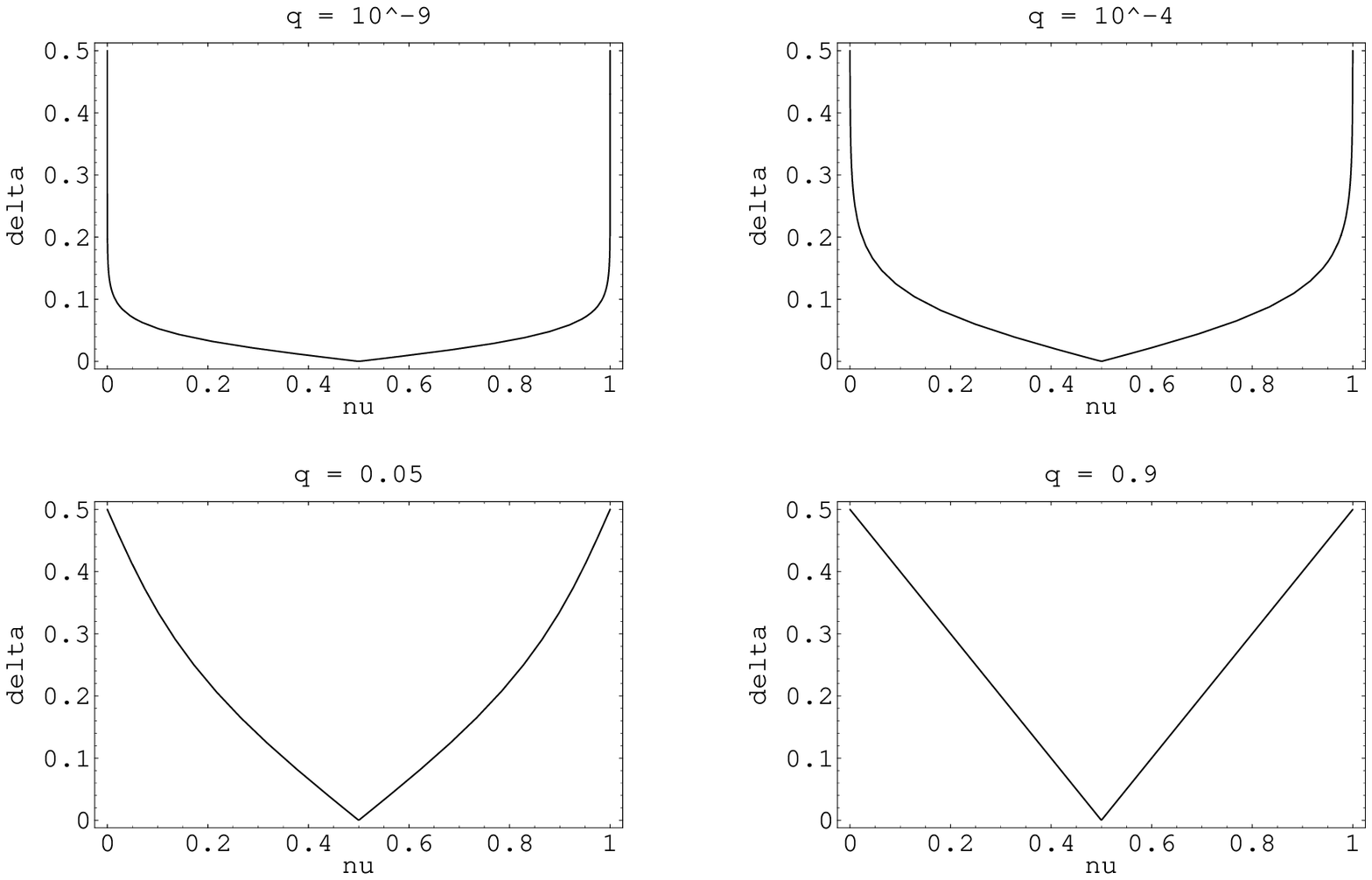}}
\parbox{14truecm}{\caption{\baselineskip=16 pt\small\label{fig:deltanu}
A plot of the function $\delta(\nu,q)$ for four different values of $q$.}
}
\end{center}
\end{figure}

\begin{proof}
The periodicity of $F_\infty$ follows directly from its definition. 
To prove (ii), define $F(\mu)$ for $0<\mu<1$ as
\be
F(\mu)
  = \sum_{k=1}^\infty
	\left[\frac{1}{1+e^{a(l-\mu)}}
	-\frac{1}{1+e^{a(l+\mu)}}\right]
    -\frac{1}{1+e^{a\mu}}
\label{otherF}\ee
Then,
\beann
F(1-\mu)&=&\sum_{l=1}^\infty\left[\frac{1}{1+e^{a(l-1+\mu)}}
-\frac{1}{1+e^{a(l+1-\mu)}}\right]-\frac{1}{1+e^{a(1-\mu)}}\\
&=&\sum_{l=1}^\infty\left[\frac{1}{1+e^{a(l+\mu)}}
-\frac{1}{1+e^{a(l-\mu)}}\right]\\
&&+\frac{1}{1+e^{a\mu}}+\frac{1}{1+e^{a(1-\mu)}}-\frac{1}{1+e^{a(1-\mu)}}\\
&=&-F(\mu)
\eeann
And clearly the remainder term
$$
\cases{
\frac{1}{2}-\{\mu\},& if $\mu \not\in \Ir$\cr
                   0 ,& if $\mu \in \Ir$}
$$
satisfies property (ii).
For the bounds, we first restrict ourselves to $\mu\in[0,1]$. 
For $\mu\geq 0$, we note that \eq{defF} implies
$$
F_\infty(\mu) \geq -\{\mu\}\geq -1.
$$
Then we use property ii) in combination with this bound
to also get the upper bound for $\mu\in[0,1]$.
$$
F_\infty(\mu)=-F_\infty(1-\mu)\leq 1
$$
Due to the peridicity property i), the upper and lower bound
are automatically extended to all real $\mu$.
The special values stated in iv) are straightforward from
\eq{defF} and \eq{otherF}.
\end{proof}
\begin{figure}[t]
\begin{center}
\resizebox{13cm}{!}{\includegraphics{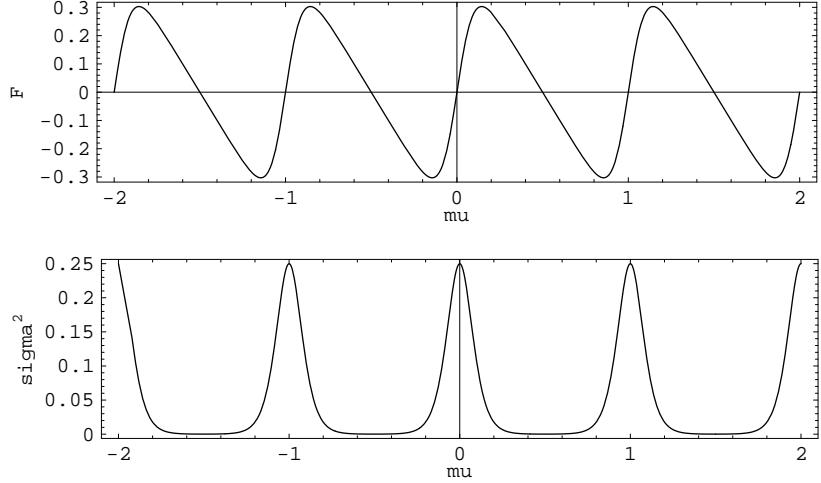}}
\parbox{14truecm}{\caption{\baselineskip=16 pt\small\label{fig:F_sigma}
A plot of the functions $F_\infty(\mu)$ and $\sigma^2(\mu)$, with $q
=e^{-10}$.}
}
\end{center}
\end{figure}
We can define the quantity $\delta(\mu)=\min(\vert\mu-\floor{\mu}\vert,
\vert 1-\mu+\floor{\mu}\vert)$, where $\floor{\mu}$ is the integer part of
$\mu$. In general, the relation between $\mu$ and $\nu$ depends
nontrivially on $q$ and the function $\delta$ can be thought as
$\delta(q,\nu)$. But for all $q$, $0<q<1$, one has $\delta(q,1/2)=0$ and
$\delta(q,0)=1/2$. See Figure \ref{fig:deltanu}.

\subsection{The variance of the number of particles in a stick}
\label{sect:var_n}

In the same way as was done above for the mean, we can compute
the variance of the number of particles in a stick in the 
grand canonical ensemble by using the standard formula
\be
\sigma^2(\mu,L) =\avgGC{\Num^2}{\Sigma,\mu} -(\avgGC{\Num}{\Sigma,\mu})^2
= \frac{1}{a^2} \frac{\partial^2}{{\partial \mu^2}} 
\log \ZGC{\Sigma}{\mu} ,
\ee 
which gives
\be
\sigma^2(\mu,L) = \frac{1}{4}\sum_{l=-L/2}^{L/2} \frac{1}{\cosh^2
(\frac{a}{2}
(l-\mu))}
\ee
Define 
\be
\sigma^2(\mu)=\lim_{L\to\infty}\sigma^2(\mu,L)
\ee
Then, the speed of convergence of this limit is bounded as follows:
\be
\vert \sigma^2(\mu) - \sigma^2(\mu,L)\vert
\leq 2\sum_{n=0}^{\infty}e^{-a(n-\mu + L/2)}=\frac{2q^{2(L/2-\mu)}}{1-q^2}
\ee
It is clear that $\sigma^2(\mu)$ is a periodic function of $\mu$ with 
period 1. It is not hard to see that $\sigma^2(\mu,L)$ is $C^\infty$ and 
attains its maximum in all integers and its minimum in the integers $+
1/2$.
It is easy to derive upper and lower bounds for $\sigma^2(\mu,L)$.
An upper bound is given by 
\be
\sigma^2(\mu,L)\leq \sum_{l=-L/2}^{L/2} e^{-\vert a(l-\mu)\vert}
\leq \sum_{l=-L/2}^{L/2} e^{-a\vert l\vert}\leq 1+
\frac{2e^{-a}}{1-e^{-a}}
\label{upper_sigma}\ee
and a lower bound  can be obtained using the crude bound
$2\cosh x\leq 2 e^{\vert x\vert }$:
\be
\sigma^2(\mu,L)\geq \frac{1}{4}\sum_{n=1}^{L} e^{-\vert an\vert}
\geq \frac{1}{4}\frac{e^{-a}-e^{-a (L+1)}}{1-e^{-a}}
\label{lower_sigma}\ee
{From} \eq{upper_sigma} and \eq{lower_sigma} we see that the limit 
$\sigma^2(\mu)$ satisfies the bounds
\be
\frac{1}{4}\frac{q^2}{1-q^2}\leq \sigma^2(\mu) \leq \frac{1+q^2}{1-q^2},
\ee
for all real $\mu$ and where we have again used the relation $e^{-a}=q^2$.

For the afficionados, one can also show that 
\be\
\lim_{q\downarrow 0}\sigma^2(\mu)=\cases{0          & if $\mu\not\in\Ir$\cr
                                         \frac{1}{4}& if $\mu\in\Ir$}					 
\ee
The interpretation is simple. When $\mu\in\Ir$, the interface (kink) in
the one-dimensional system is located at a lattice site, which is occupied
by a particle with probability 1/2. Clearly, the variance of the particle
number is them $1/4$. However, for $\mu\not\in\Ir$, the kink is centered
at a position not belonging to the lattice and the state converges, as
$q\downarrow 0$, to a deterministic configuration with zero variance
for the particle number.

\subsection{Estimating $\avgGC{q^{2|\Num - \avg{\Num}{}|}}{\Sigma,\mu}$}
\label{sect:avgexp}
We begin with the obvious fact
\be
q^{2|\Num - \avg{\Num}{}|} 
  \leq q^{2\Num - 2\avg{\Num}{}} + q^{2\avg{\Num}{} - 2\Num}
\ee
from which it follows that
\be
\avgGC{q^{2|\Num - \avg{\Num}{}|}}{\Sigma,\mu}
  \leq q^{-2\avg{\Num}{}} \avgGC{q^{2\Num}}{\Sigma,\mu}
  + q^{2\avg{\Num}{}} \avgGC{q^{-2\Num}}{\Sigma,\mu}.
\ee
Now, we observe
\be
\label{qexp:avg}
\avg{q^{2\Num}}{\Sigma,\mu} 
  = \frac{\sum_{n=0}^{L+1} q^{2n} q^{-2 \mu n} Z(\Sigma,n)}
{Z^{GC}(\Sigma,\mu)}
  = \frac{Z^{GC}(\Sigma,\mu-1)}{Z^{GC}(\Sigma,\mu)} .
\ee
Since
\be
  Z^{GC}(\Sigma,\mu) = \prod_{l=-L/2}{L/2} (1 + q^{2(l - \mu)})
\ee
equation \eq{qexp:avg} leads us to conclude
\be
\avg{q^{2\Num}}{\Sigma,\mu}
  = \frac{1 + q^{2(L/2+1-\mu)}}{1+q^{-2(L/2+\mu)}} 
  \leq 2 q^{2(L/2+\mu)}.
\ee
Similarly,
\be
\avg{q^{-2\Num}}{\Sigma,\mu}
  = \frac{1 + q^{-2(L/2+1+\mu)}}{1 + q^{2(L/2 - \mu)}}
  \leq 2 q^{-2(L/2+1+\mu)}.  
\ee
Using the results of section \ref{sect:navg}, we then have
\be
\avgGC{q^{2|\Num-\avg{\Num}{}|}}{\Sigma,\mu}
	\leq 4 q^{-1-|F_L(\mu)|}
	\leq 4 q^{-2}.
\ee
If we wish to calculate $\avgGC{q^{2|\Num-\avg{\Num}{}|}}{\Lambda,\mu}$,
where $\Lambda$ is comprised of $A$ sticks, then nothing changes except 
that each estimate is raised to the power $A$.
Thus, $\avgGC{q^{2|\Num-\avg{\Num}{}|}}{\Lambda,\mu} \leq 2^{A+1} q^{-2A}$.

\section*{Acknowledgements}

O.B. was supported by Fapesp under grant 97/14430-2. B.N. was partially
supported by the National Science Foundation under grant \# DMS-9706599.

\end{document}